\documentclass[twocolumn,aps,superscriptaddress,showpacs,nofootinbib,floatfix]{revtex4}
\usepackage{epsfig,bm,feynmf}
\usepackage{graphics}
\usepackage{amsmath}
\usepackage[normalem]{ulem}  
\usepackage[dvips]{color} 

\renewcommand\sout{\bgroup \color{red} \ULdepth=-.5ex \ULset}
\begin{document}


\title{Charm production in Pb+Pb collisions at the Large Hadron Collider energy}


\author{Taesoo Song}\email{song@fias.uni-frankfurt.de}
\affiliation{Institute for Theoretical Physics, Johann Wolfgang Goethe Universit\"{a}t, Frankfurt am Main, Germany}
\affiliation{Frankfurt Institute for Advanced Studies, Johann Wolfgang Goethe Universit\"{a}t, Frankfurt am Main, Germany}

\author{Hamza Berrehrah}
\affiliation{Institute for Theoretical Physics, Johann Wolfgang Goethe Universit\"{a}t, Frankfurt am Main, Germany}
\affiliation{Frankfurt Institute for Advanced Studies, Johann Wolfgang Goethe Universit\"{a}t, Frankfurt am Main, Germany}

\author{Daniel Cabrera}
\affiliation{Institute for Theoretical Physics, Johann Wolfgang Goethe Universit\"{a}t, Frankfurt am Main, Germany}
\affiliation{Frankfurt Institute for Advanced Studies, Johann Wolfgang Goethe Universit\"{a}t, Frankfurt am Main, Germany}

\author{Wolfgang Cassing}
\affiliation{Institut f\"{u}r Theoretische Physik, Universit\"{a}t Gie$\beta$en, Germany}

\author{Elena Bratkovskaya}
\affiliation{Institute for Theoretical Physics, Johann Wolfgang Goethe Universit\"{a}t, Frankfurt am Main, Germany}
\affiliation{Frankfurt Institute for Advanced Studies, Johann Wolfgang Goethe Universit\"{a}t, Frankfurt am Main, Germany}


\begin{abstract}
We study charm production in Pb+Pb collisions at $\sqrt{s_{\rm
NN}}=$2.76 TeV in the Parton-Hadron-String-Dynamics transport approach and the charm
dynamics in the partonic and hadronic medium. The charm quarks are
produced through initial binary nucleon-nucleon collisions by using
the PYTHIA event generator taking into account the (anti-)shadowing
incorporated in the EPS09 package. The produced charm quarks
interact with off-shell massive partons in the quark-gluon plasma and are
hadronized into $D$ mesons through coalescence or fragmentation
close to the critical energy density, and then interact with hadrons
in the final hadronic stage with scattering cross sections
calculated in an effective Lagrangian approach with heavy-quark
spin symmetry. The PHSD results show a reasonable $R_{\rm AA}$ and
elliptic flow of $D$ mesons in comparison to the experimental data
for Pb+Pb collisions at $\sqrt{s_{NN}}$ = 2.76 TeV from the ALICE
Collaboration. We also study the effect of
temperature-dependent off-shell charm quarks in relativistic
heavy-ion collisions. We find that the scattering cross sections
are only moderately affected by off-shell charm degrees of freedom.
However, the position of the peak of $R_{\rm AA}$ for $D$ mesons
depends on the strength of the scalar partonic forces which also
have an impact on the $D$ meson elliptic flow. The comparison with
experimental data on the $R_{\rm AA}$ suggests that the repulsive
force is weaker for off-shell charm quarks as compared to that  for
light quarks. Furthermore, the effects from radiative charm energy
loss appear to be low compared to the collisional energy loss up to
transverse momenta of $\sim$ 15 GeV/c.
\end{abstract}


\maketitle

\section{Introduction}

The strong interaction, which mediates the energy-momentum exchange
between hadrons as well as partons, is described by the Quantum
Chromo Dynamics (QCD). The characteristic features of QCD are the
asymptotic freedom at short distance and the confinement at long
distance. Due to these features of QCD, the partons behave as free
particle at short distance but are confined inside hadrons on
distances of the order $\sim$ 1 fm. With increasing temperature or
nuclear density the hadrons overlap in space, and the partons --
confined before in a single hadron -- now can freely move for distances that
are large compared to the hadron size. The phenomenon  is called
deconfinement or the phase transition to a quark-gluon plasma (QGP).

Relativistic heavy-ion collisions are the experiments to realize such
extreme conditions. Since the hot and dense matter produced in
relativistic heavy-ion collisions disappears on time scales of a couple of fm/c, it is
a big challenge to investigate its properties. One can obtain
information on the system by measuring bulk particles,
electromagnetic probes such as direct photons or lepton pairs or hard particles.
The hard particles
are normally represented by jets and heavy flavors. The former are light particles
with large transverse momentum and neighbours in a momentum cone, while the
latter represent heavy particles
which have charm or bottom flavor. Since the production of hard
particles requires large energy-momentum transfer, it takes place early in
relativistic heavy-ion collisions, and can be
described by perturbative QCD (pQCD).

The produced hard particle interacts with the hot dense matter
by exchanging energy and momentum. For example, a hard particle
with large transverse momentum (relative to the bulk matter) loses part
of its energy while passing through the
medium. This results in a suppression of the $R_{\rm AA}$ at high
transverse momentum, which is the ratio of the measured particle number
in heavy-ion collisions to the expected number in the absence of
nuclear or partonic matter. With increasing strength of the interaction
of a hard particle with
the medium the ratio $R_{\rm AA}$ becomes more suppressed at high
transverse momentum.

It has been  naively expected that the $R_{\rm AA}$ of heavy flavor mesons
is less suppressed as compared to that of light hadrons for two reasons:
Firstly, the scattering cross section of a heavy quark is smaller in pQCD than
that of  a gluon which produces e.g. a light-hadron jet.
Secondly, the gluon radiation from heavy quarks is suppressed due to so-called
dead-cone effect~\cite{Dokshitzer:2001zm}.
However, experimental data show that the suppression of heavy-flavor mesons is
comparable to that of light hadrons \cite{ALICE:2012ab}.
Also the elliptic flow of heavy-flavor hadrons is not small compared to that of
light hadrons~\cite{Abelev:2013lca} and of comparable size. This sets up a puzzle
for heavy-flavor production in relativistic heavy-ion collisions.

There have been various theoretical studies on the heavy-quark diffusion in relativistic heavy-ion collisions.
Most of them are based on the Boltzmann equation~\cite{Molnar:2006ci,Zhang:2005ni,Linnyk:2008hp,Uphoff:2011ad,Uphoff:2012gb,Gossiaux:2010yx,Nahrgang:2013saa};  an  alternative way is to solve the Langevin equation~\cite{Moore:2004tg,He:2011qa,He:2012df,Cao:2011et,Cao:2015eek}
for the charm dynamics as an approximation to the  Boltzmann equation. The latter again is closely linked to
the Fokker-Plank equation connecting drift and diffusion by the Einstein relation at fixed temperature.

Since the heavy-flavor interaction is intimately related to the dynamics of the partonic or hadronic bulk matter, a proper description of the relativistic heavy-ion collisions is essential.
The models describing the heavy-ion dynamics are classified into macroscopic and microscopic ones.
Hydrodynamic simulations are a macroscopic description which assume local thermal
equilibrium and numerically solve intensive thermal quantities as functions of space and time by choosing a proper Equation-of-State (EoS).
There have been several attempts to include off-equilibrium effects by
introducing viscosity or  anisotropic momentum distributions of the hydro fluid.
On the other hand, microscopic approaches are based on the Boltzmann equation or some extensions of it.
For a review on the different approaches we refer the reader to Ref. \cite{Andronic:2015wma}.

The Parton-Hadron-String Dynamics (PHSD) approach, which we use in this study, differs in
several aspects from the conventional Boltzmann-type approaches~\cite{Cassing:2009vt}.
First of all, the degrees-of-freedom for the QGP phase are massive strongly-interacting quasi-particles.
The masses of the dynamical quark and gluon in the QGP are distributed according to spectral functions whose pole positions and widths, respectively, are defined by the real and imaginary parts of their self-energies.
The latter are defined in the dynamical quasiparticle model (DQPM) in which the strong coupling and the self-energies  are fitted to lattice QCD results. Due to the finite spectral width, the spectral function has time-like as well as space-like parts. The time-like partons propagate in space-time within the light-cone while the space-like components are attributed to a scalar potential energy density~\cite{Cassing:2009vt}.
The gradient of the potential energy density with respect to the scalar density generates a repulsive force in relativistic heavy-ion collisions and plays an essential role in reproducing experimental flow data and transverse momentum spectra.
We recall that the PHSD approach has successfully described numerous experimental data in relativistic heavy-ion collisions from the Super Proton Synchrotron (SPS) to Large Hadron Collider (LHC) energies~\cite{Cassing:2009vt,PHSDrhic,Volo,Linnyk};
a review on bulk and electromagnetic properties of relativistic heavy-ion reactions within PHSD can be found in Ref.~\cite{PPNP16}.

Recently, explicit  charm production has been implemented in the PHSD~\cite{Song:2015sfa}.
The initial charm and anticharm quarks are produced by using the PYTHIA event generator.
In the QGP they interact with off-shell partons and finally are hadronized into $D$ mesons either through fragmentation or coalescence if the energy density is close to the critical energy density for the crossover transition ($\sim$ 0.5 GeV/fm$^3$).
The hadronized $D$ mesons then interact with light hadrons and finally freeze out.
The PHSD approach has been applied for charm production in Au+Au collisions at $\sqrt{s_{\rm NN}}=$200 GeV, and the results on the $R_{\rm AA}$ as well as the $v_2$ of $D$ mesons are in reasonable agreement with the experimental data from the STAR collaboration~\cite{Adamczyk:2014uip,Tlusty:2012ix} as demonstrated in Ref. \cite{Song:2015sfa}.

In this study we extend our previous work and apply the PHSD to  charm production in Pb+Pb collisions at $\sqrt{s_{\rm NN}}=$2.76 TeV. In this way we can
examine the validity and consistency of the PHSD approach in describing charm production in relativistic heavy-ion collisions in a wide range of collision energies in connection with our previous study~\cite{Song:2015sfa} and in connection with the
dynamics of light-flavor hadrons \cite{PPNP16}.
We also study the effects of parton (anti-)shadowing and of off-shell charm and
anticharm quarks on the charm production and dynamics in relativistic heavy-ion collisions at the LHC.

This paper is organized as follows:
The charm production in initial nucleon-nucleon binary collisions is described in Sec.~\ref{initial} and the
results are compared with those from  Fixed-Order Next-to-Leading Logarithm (FONLL) calculations and the experimental data in p+p collisions.
In Sec.~\ref{shadowing}, we explain how the (anti-)shadowing effect is implemented in PHSD which had been discarded
in Ref. \cite{Song:2015sfa} at RHIC energies.
We then describe the partonic and hadronic interactions of charm as well as its hadronization in Sec.~\ref{interaction}.
Finally, the nuclear modification and elliptic flow of $D$ mesons from the PHSD are shown in Sec.~\ref{results} and compared with the experimental data from the ALICE collaboration.
Sec.~\ref{summary} gives a summary of the present work.

\section{Initial charm quark production}\label{initial}

\begin{figure}[h]
\centerline{
\includegraphics[width=9.5 cm]{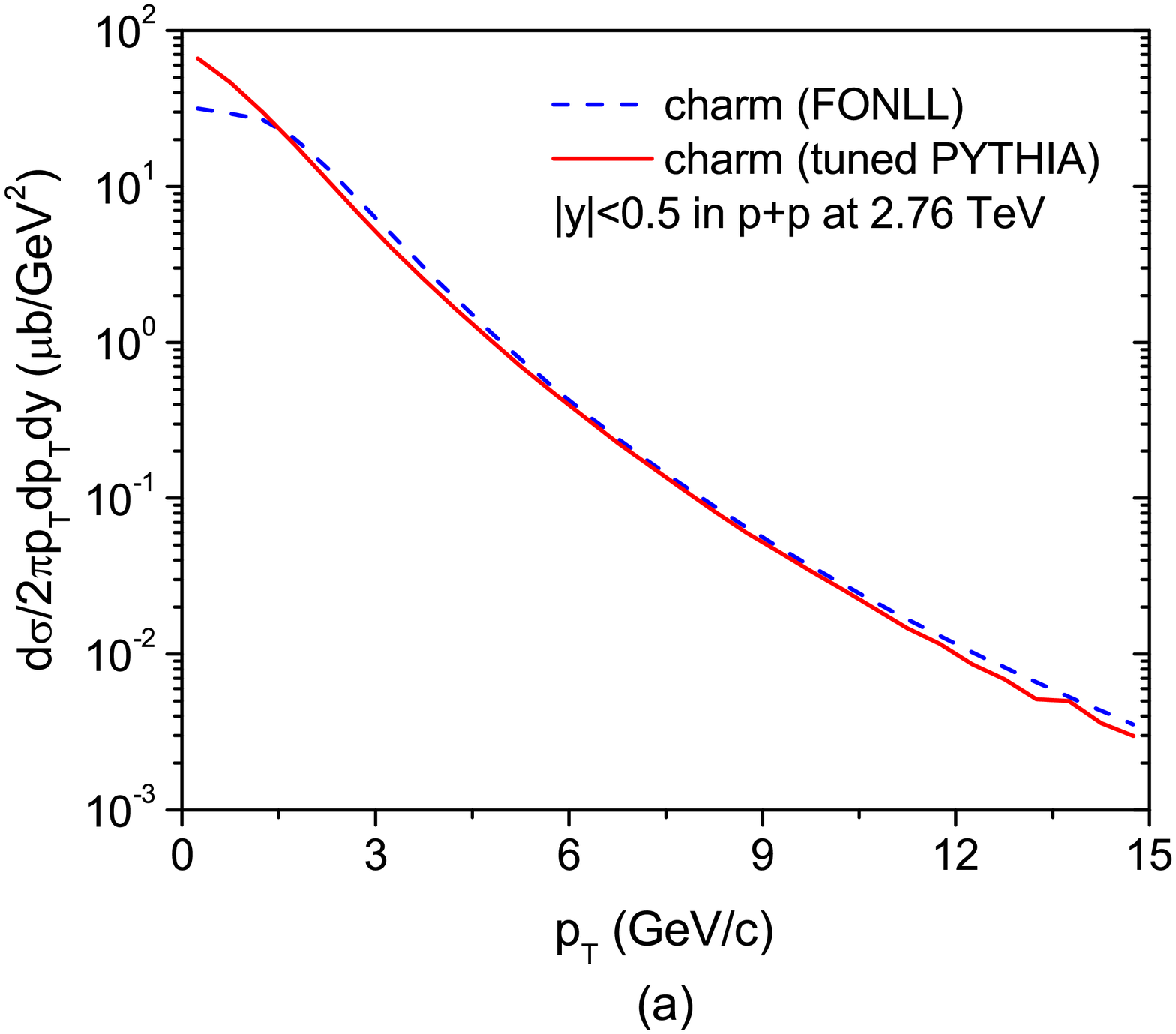}}
\centerline{
\includegraphics[width=9.5 cm]{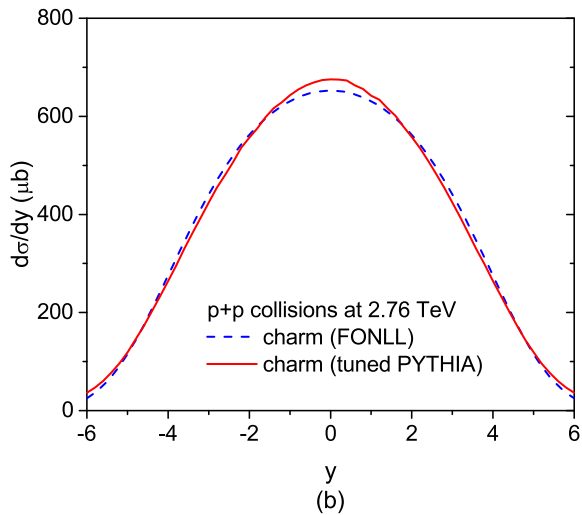}}
\caption{(Color online) Transverse momentum (a) and rapidity (b)
distributions of charm quarks in p+p collisions at $\sqrt{s_{\rm NN}}=$2.76
TeV from FONLL (dashed lines) and the tuned PYTHIA event generator (solid lines).}
\label{pp1}
\end{figure}

We generate initial charm and anticharm quark pairs by using the PYTHIA event
generator~\cite{Sjostrand:2006za}.
In order to reproduce the differential cross sections for charm quark production
from the Fixed-Order Next-to-Leading Logarithm (FONLL)
calculations~\cite{Cacciari:2012ny}, the rapidity distribution of the charm quark is
increased by 9 \%.
In Fig.~\ref{pp1} we compare the transverse momentum spectrum and the rapidity
distribution of charm quarks in p+p collisions at $\sqrt{s_{\rm NN}}=$2.76 TeV
from the FONLL and those from the tuned PYTHIA event generator.
We note that the differential cross section from the FONLL is rescaled for comparison.

The produced charm and anticharm quarks in p+p collisions hadronize by emitting soft
gluons. The probabilities for a charm quark to hadronize into
$D^+,~D^0,~D_s^+$, and $D^{*+}$ are, respectively, taken to be 0.226, 0.557, 0.101,
and 0.238 from the combined $e^+e^-$ data with $D^*$ decay into $D^+$ and $D^0$ being
included~\cite{Gladilin:1999pj,Chekanov:2007ch,Abelev:2012vra}.
The fraction of $D^{*0}$ is given by multiplying to the fraction of $D^{*+}$ the ratio
of neutral to charged $D$ meson production rate, $R_{u/d}$, which is taken to
be 1.09~\cite{Chekanov:2007ch}.
The three-momentum of a hadronized $D$ meson is given by the fragmentation
function~\cite{Peterson:1982ak},
\begin{eqnarray}
D_Q^H(z)\sim \frac{1}{z[1-1/z-\epsilon_Q/(1-z)]^2},
\end{eqnarray}
where $z$ is the momentum fraction of the hadron $H$ fragmented from
the heavy quark $Q$ while $\epsilon_Q$ is a fit parameter which is
taken to be $\epsilon_Q$ = 0.01 as in our previous study~\cite{Song:2015sfa}.
The energy of the $D$ meson is adjusted to be on mass-shell.

\begin{figure}[h]
\centerline{
\includegraphics[width=9.5 cm]{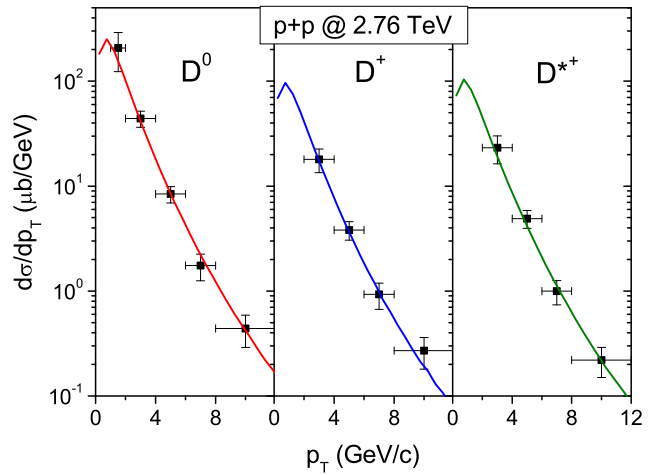}}
\caption{(Color online) Differential cross sections for $D^0$, $D^+$,
and $D^{*+}$ production {at mid-rapidity $(|y|<0.5)$} in p+p collisions
at $\sqrt{s_{\rm NN}}=$2.76 TeV from the ALICE collaboration~\cite{Abelev:2012vra}
compared with those from the tuned PYTHIA event generator and the fragmentation function
of Peterson~\cite{Peterson:1982ak} (solid lines).}
\label{pp2}
\end{figure}

Fig.~\ref{pp2} shows the differential cross sections of $D^0$, $D^+$, and
$D^{*+}$ mesons after the charm fragmentation in p+p collisions at
$\sqrt{s_{\rm NN}}=$2.76 TeV in comparison with the experimental data from the
ALICE collaboration~\cite{Abelev:2012vra}.
The agreement with the experimental data is sufficiently good.

\section{(anti-)shadowing effects}\label{shadowing}

In pQCD, a charm quark pair is produced through parton scattering.
The partonic scattering cross section for charm production is then weighted by
parton distribution functions of the nucleon in order to calculate the
production cross section in nucleon-nucleon collisions:
\begin{eqnarray}
\sigma_{c\bar{c}}^{NN}(s)=\sum_{i,j}\int dx_1dx_2 f_i^N(x_1,Q)f_j^N(x_2,Q)\nonumber\\
\times\sigma_{c\bar{c}}^{ij}(x_1x_2s,Q),
\label{factorize}
\end{eqnarray}
where $f_i^N(x,Q)$ is the distribution function of the parton $i$ with the
energy-momentum fraction $x$ in the nucleon at scale $Q$. The
momentum fractions
$x_1$ and $x_2$ are calculated from the transverse mass ($M_T$) and the
rapidity ($y$) of the final-state particles by
\begin{eqnarray}
x_1&=&\frac{M_T}{E_{\rm cm}}e^{y},\nonumber\\
x_2&=&\frac{M_T}{E_{\rm cm}}e^{-y},
\label{x1x2}
\end{eqnarray}
where $E_{\rm cm}$ is the nucleon-nucleon collision energy in the center-of-mass frame.
We recall that a nucleon is occupied by valence quarks at large $x$ and dominantly by gluons at
small $x$.
Since charm-quark pair production requires a large energy-momentum transfer,
partons with large $x$ dominantly contribute to the production.
However, with increasing  collision energy, partons with small $x$ are more and
more involved in charm pair production.
As a result, gluon fusion becomes more important for charm production than quark and
antiquark annihilation for such high energy collisions as at the LHC.

\begin{figure}[h]
\centerline{
\includegraphics[width=10 cm]{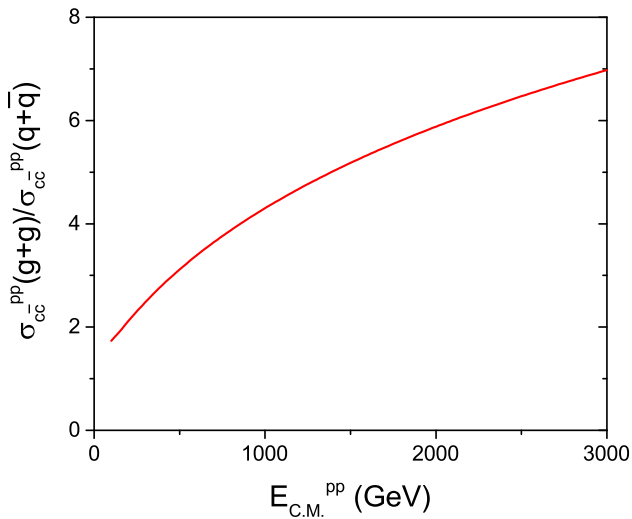}}
\caption{(Color online)
{Ratio of the cross section for charm production in the channel $g+g\rightarrow c+\bar{c}$ to that
in $q+\bar{q}\rightarrow c+\bar{c}$ in p+p collisions as a function of collision
energy.} The partonic cross sections are calculated up to LO in
pQCD~\cite{Combridge:1978kx}, and the CTEQ6M parton distribution
function~\cite{Pumplin:2002vw} is used in Eq.~(\ref{factorize}).}
\label{pp-LO}
\end{figure}

Fig.~\ref{pp-LO}
{shows the ratio of the cross section for charm production in the channel
$g+g\rightarrow c+\bar{c}$ to that in $q+\bar{q}\rightarrow c+\bar{c}$ in p+p
collisions as a function of the collision energy.}
The partonic cross sections are calculated up to Leading-Order (LO) in pQCD~\cite{Combridge:1978kx}, and the CTEQ6M parton distribution function~\cite{Pumplin:2002vw}  in Eq.~(\ref{factorize}).
Fig.~\ref{pp-LO} shows that gluon fusion is much more important than quark and
antiquark annihilation for charm pair production in high-energy  collisions.
Therefore, we assume in this study that all charm pairs are produced through gluon fusion.

We recall that the parton distribution function (PDF) is modified in a nucleus to
\begin{eqnarray}
f_i^{N^*}(x,Q)=R_i^A(x,Q)f_i^N(x,Q),
\label{shadow}
\end{eqnarray}
where $N^*$ indicates the nucleon in nucleus $A$, and $R_i^A(x,Q)$ is the ratio of
the PDF of $N^*$ to that of a free nucleon.
The ratio $R_i^A(x,Q)$ of a heavy nucleus $A$, which is lower than 1 at small $x$,
increases with increasing $x$, and is slightly larger than 1 from a certain value of $x$.
The region $R_i^A(x,Q) < 1$ is called shadowing, and the regime $R_i^A(x,Q) > 1$ is called
antishadowing.
The EPS09 package -- used in this study -- parameterizes $R_i^A(x,Q)$
from a fit of the parameters to the experimental data from deep inelastic
$l+$A scattering, Drell-Yan dilepton production in p+A collisions,
and inclusive pion production in d+Au and p+p collisions at RHIC~\cite{Eskola:2009uj}.

Substituting Eq.~(\ref{shadow}) into Eq.~(\ref{factorize}), the cross section
for charm production is modified to
\begin{eqnarray}
\sigma_{c\bar{c}}^{N^*N^*}(s)=\sum_{i,j}\int dx_1dx_2 R_i^A(x_1,Q)R_j^A(x_2,Q) \nonumber\\
\times f_i^N(x_1,Q)f_j^N(x_2,Q)\sigma_{c\bar{c}}^{ij}(x_1x_2s,Q).
\label{factorize2}
\end{eqnarray}

In order to include the (anti-)shadowing effect in the PHSD, we take the following steps:
First, the energy-momentum fractions $x_1$ and $x_2$ are calculated from the
transverse mass and rapidity of the charm quark pair, which is generated by
PYTHIA by using Eq.~(\ref{x1x2}).
Second, $R_i^A(x_1,Q)$ and $R_j^A(x_2,Q)$ for $i=j={\rm gluon}$ are obtained from
the EPS09 package. The scale $Q$ is taken to be the average of the transverse mass
of the charm and that of the anticharm.
Third, we introduce a maximum value of $R_i^A(x_1,Q)R_j^A(x_2,Q)$, for example, 10.
If a random number is larger than the ratio of $R_i^A(x_1,Q)R_j^A(x_2,Q)$ to the
maximum value, the produced charm quark pair is discarded, and a new charm quark pair
is generated by PYTHIA.
These steps are repeated until the random number is smaller than the ratio.

The (anti-)shadowing effect is expected to depend on the impact parameter in heavy-ion
collisions such that it is strong in central collisions and weak in peripheral collisions.
We assume that the (anti-)shadowing effect is proportional to the thickness of the
nucleus,
\begin{eqnarray}
T_A(r_\bot)=\frac{3N_A}{2\pi R_A^2}\sqrt{1-\frac{r_\bot^2}{R_A^2}},
\end{eqnarray}
where $N_A$ and $R_A$ are the mass number and the radius of nucleus $A$ and $r_\bot$
is the transverse distance from the center of the nucleus.
For $R_i^A(x_1,Q)$, the averaged ratio over  impact parameter, we find
\begin{eqnarray}
R_i^A(r_\bot,x,Q)=\frac{4}{3}\sqrt{1-\frac{r_\bot^2}{R_A^2}}~R_i^A(x,Q),
\label{impact}
\end{eqnarray}
which satisfies
\begin{eqnarray}
R_i^A(x,Q)=\frac{2\pi}{N_A}\int dr_\bot r_\bot^2T_A(r_\bot)R_i^A(r_\bot,x,Q).
\end{eqnarray}

The (anti-)shadowing affects the total cross section for charm production as well as
the $\rm p_T$ spectrum of produced charm.
Before taking the above steps, therefore, we precalculate the total cross section for each centrality from
\begin{eqnarray}
\frac{\sigma_{c\bar{c}}^{N^*N^*}(s)}{\sigma_{c\bar{c}}^{NN}(s)}= \frac{1}{n}\sum_{i=1}^n R_g^{Pb}(r_{\bot i}^A,x_{1i},Q_i)R_g^{Pb}(r_{\bot i}^B,x_{2i},Q_i),
\end{eqnarray}
where $n$ is the number of PYTHIA events for charm production in heavy-ion collisions, and $r_{\bot i}^A$ and $r_{\bot i}^B$, respectively, the transverse positions of the production from the center of nucleus A and nucleus B;
$x_{1i}$ and $x_{2i}$ are calculated in each PYTHIA event by using Eq.~(\ref{x1x2}).

\begin{figure}[h]
\centerline{
\includegraphics[width=10 cm]{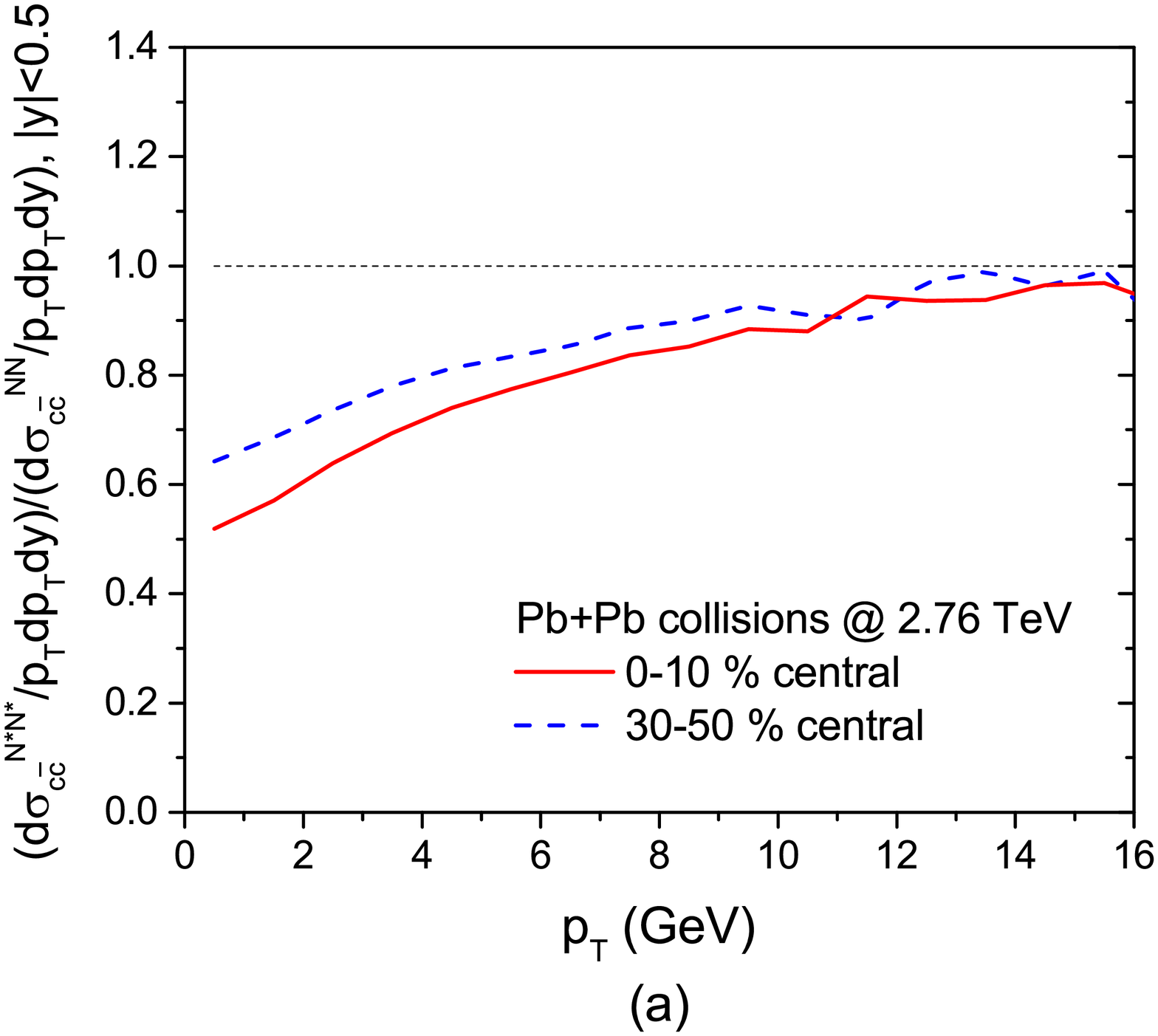}}
\centerline{
\includegraphics[width=10 cm]{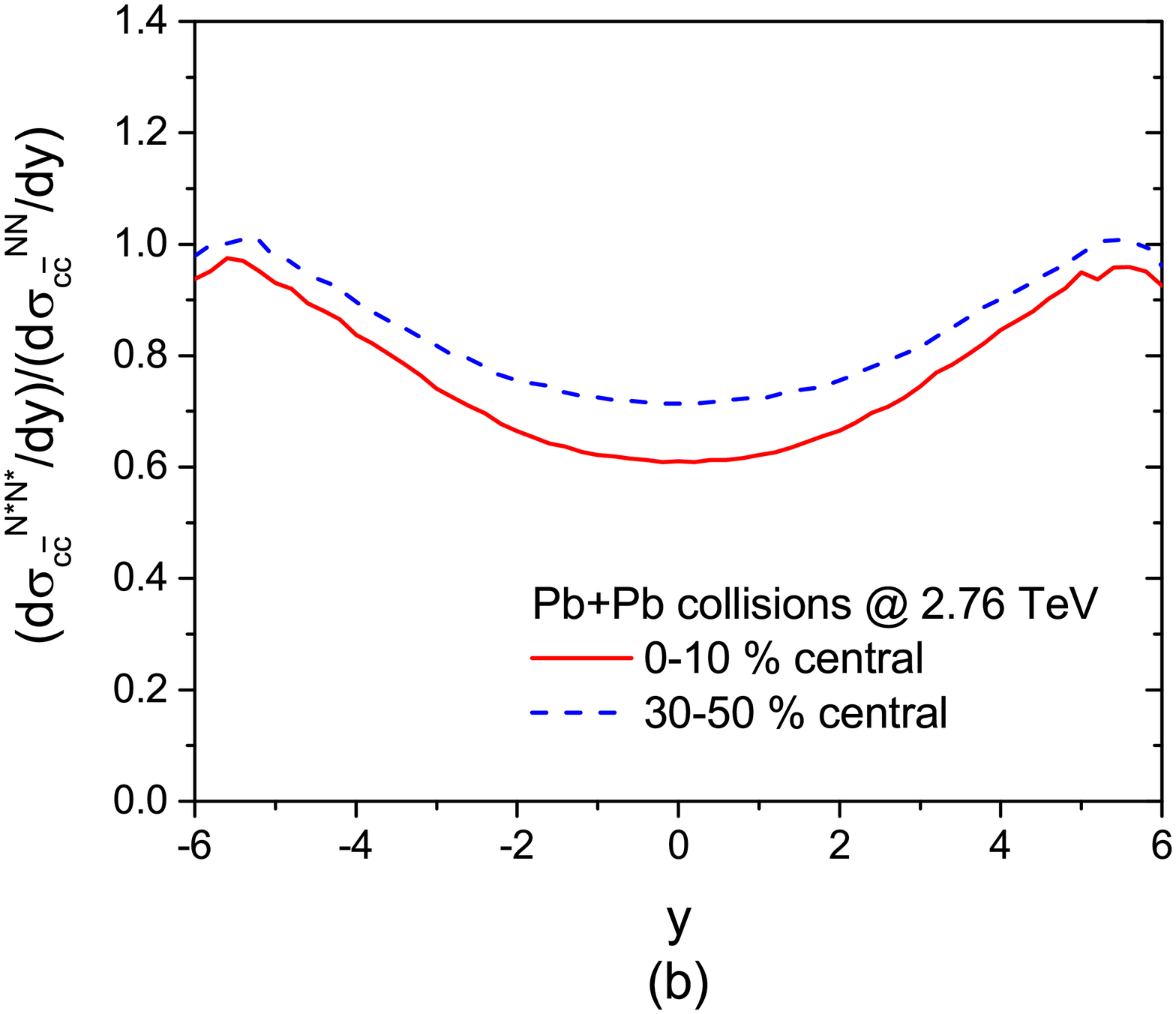}}
\caption{(Color online) The ratio of differential cross section for charm production in heavy-ion collisions to that in p+p collisions as functions
of transverse momentum $p_T$ for $|y|<0.5$ (a) and of rapidity (b) due to
the (anti-)shadowing in 0-10 \% and 30-50 \% central Pb+Pb collisions at
$\sqrt{s_{\rm NN}}=$2.76 TeV.}
\label{shadow0}
\end{figure}

Fig.~\ref{shadow0} shows the ratio of differential cross section for charm production in heavy-ion collisions to that in p+p collisions as functions
of transverse momentum $p_T$ for $|y|<0.5$ (a) and of rapidity (b) due to (anti-)shadowing
in 0-10 \% and 30-50 \% central Pb+Pb collisions at $\sqrt{s_{\rm NN}}=$2.76 TeV.
The production of charm and anticharm quarks is suppressed near mid-rapidity and
at low transverse momentum which correspond to small $x$ in the PDF; the suppression
is larger in central collisions.
Our results of shadowing effect are consistent with those from Ref.~\cite{Cao:2015hia}, because the same EPS09 package is used. However, we take into account the centrality dependence by using Eq.~(\ref{impact}).
We note that charm production is suppressed by $\sim $30 \% in 0-10 \% central collisions
and by $\sim $ 20 \% in 30-50 \% central collisions.

\section{Charm interactions in the QGP}\label{interaction}
\subsection{partonic interactions}

In PHSD, baryon-baryon and baryon-meson collisions at high-energy  produce strings.
If the local energy density is above the critical energy density ($\sim$ 0.5
GeV/fm$^3$),
the strings melt into quarks and antiquarks with masses  determined by the
temperature-dependent spectral functions from the DQPM~\cite{Cassing:2008nn}.
Massive gluons then are formed through flavor-neutral quark and antiquark fusion.
In contrast to conventional elastic scattering, off-shell partons change their mass
after the elastic scattering according to the local 'temperature' (energy density) in the local cell
where the scattering happens.
This automatically updates the parton mass distribution when the  hot and dense  matter
expands, i.e. the local 'temperature' decreases with time.

We note that the spectral function of charm or anticharm quarks cannot be fitted from
lattice QCD data
because the contribution from charm or anticharm quarks to the lattice entropy is small in the temperature region of interest.
Therefore, we adopt two scenarios:
In the first scenario, any thermal effect on the charm quark mass is completely
ignored, and the charm quark mass is 1.5 GeV regardless of temperature.
In the second scenario, the temperature dependence of the pole position and width
of the charm quark spectral function is exactly the same as that of light quarks
apart from an increase of the charm-quark pole mass by 1.0 GeV.
In the latter case, the average charm quark mass initially is $\sim$ 1.5 GeV,
i.e. the same as in the former case.
However, the charm quark, which scatters in the QGP, changes its mass according
to its thermal spectral function.

Different from the usual treatment of heavy-quark scattering using the leading-order
QCD perturbation theory (pQCD)~\cite{Combridge:1978kx,pQCD} or the inclusion of
nonperturbative features in thermal perturbation theory, denoted as the hard
thermal loop (HTL) approach~\cite{HTL}, we consider all the effects of the
nonperturbative nature of the strongly interacting quark-gluon plasma (sQGP)
constituents, i.e. the large coupling, the multiple scattering, etc. To do so,
we refrain from a fixed-order thermal loop calculation relying on perturbative
self-energies (calculated in the limit of infinite temperature) to fix the in-medium
masses of the quarks and gluons and pursue instead a more phenomenological approach.
The multiple strong interactions of quarks and gluons in the sQGP are encoded in
their effective propagators with broad spectral functions. The effective
propagators, which can be interpreted as resummed propagators in a hot and dense
QCD environment, have been extracted from lattice data in the scope of the
DQPM~\cite{Cassing:2009vt,Berrehrah:2013mua}.

The leading order processes for the scattering of a heavy quark off a light quark
and gluon are $q Q \rightarrow q Q$ and $g Q \rightarrow g Q$. In
Refs.~\cite{Berrehrah:2013mua,Berrehrah:2014kba,Berrehrah:2015ywa} we have
calculated the transition matrix elements for these  processes, considering
the effects of finite masses and widths of the different partons as well as the
scattering angle, temperature and energy dependencies of the corresponding
scattering cross sections. In this section, we highlight the differences between
the two scenarios mentioned above (charm quark mass fixed at 1.5 GeV and
off-shell charm quark masses)
on the scattering angle, temperature and energy dependence of the cross sections
as well as the charm transport coefficients.

\begin{figure}[h]
\centerline{
\hspace{-0.8 cm}\includegraphics[width=8 cm]{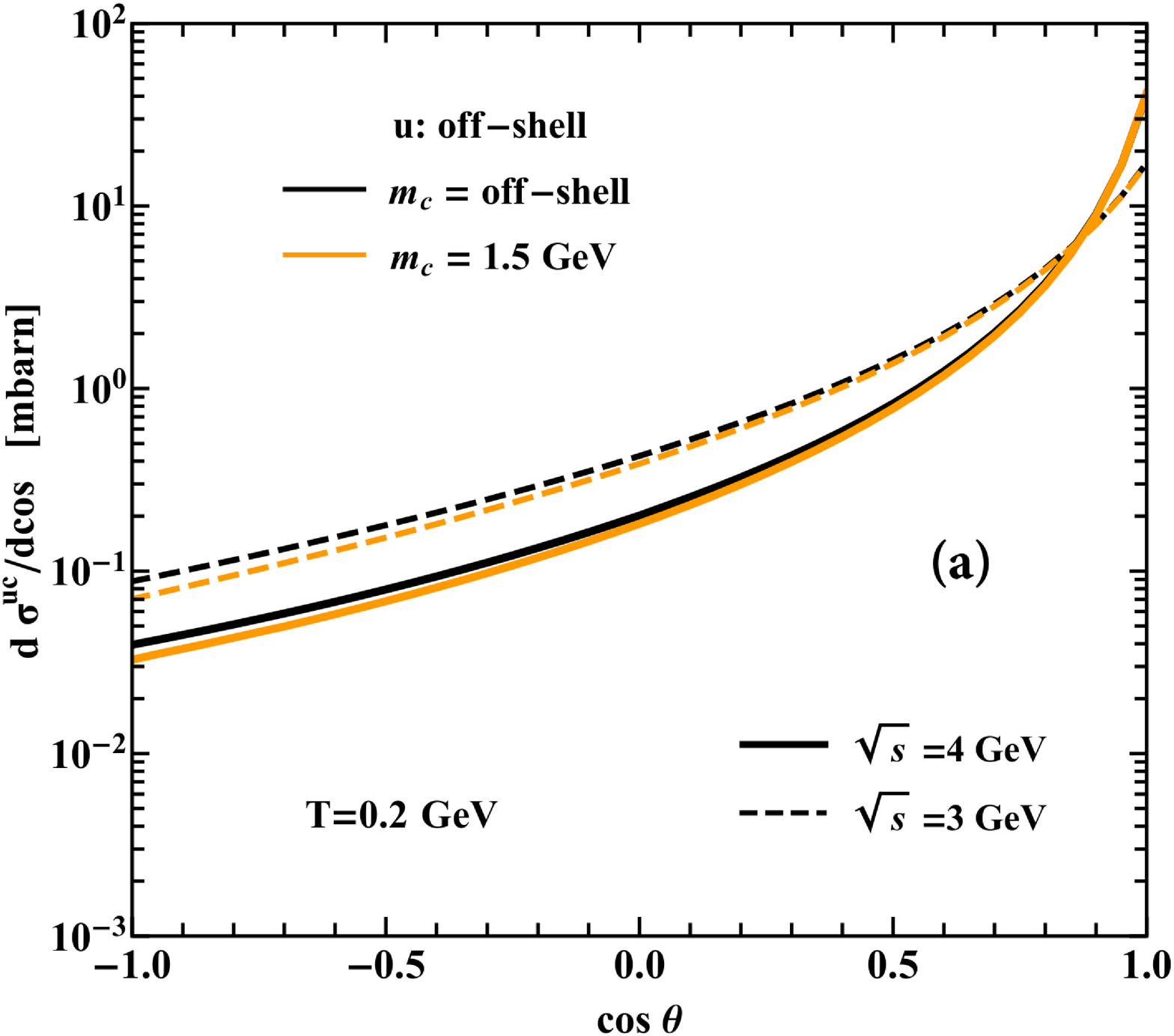}}
\centerline{
\includegraphics[width=8.2 cm]{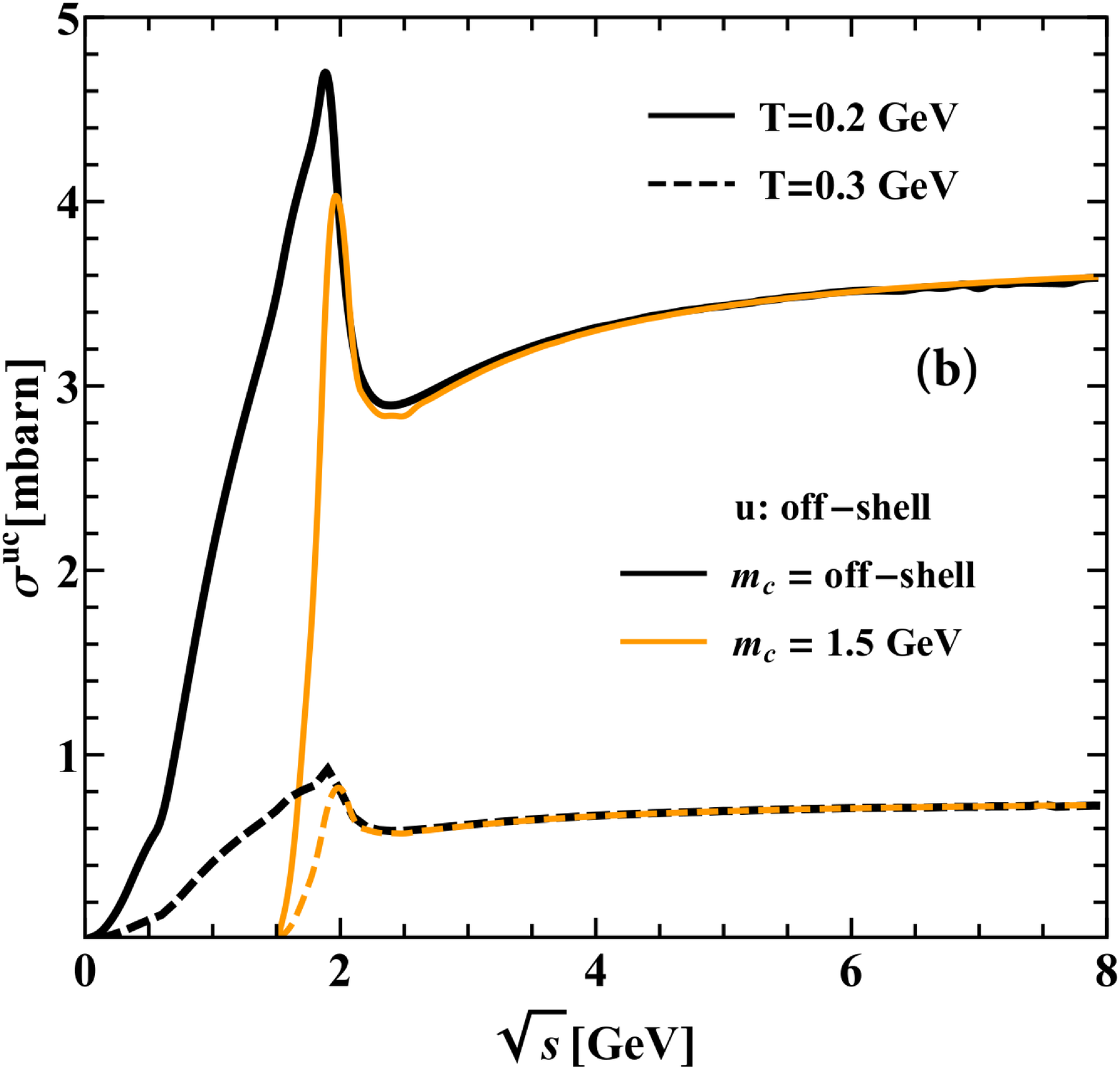}}
\caption{(Color online) The $q Q$ differential elastic cross sections as a function
of scattering angle for invariant energy $\sqrt{s}=~$3 and 4 GeV (a),
and the $qQ$ total elastic cross section as a function of the center-of-mass energy (b).
The temperature of the QGP medium is taken to be $T$=0.2 GeV in the upper panel (a),
and $T$=0.2 and $T$=0.3 GeV in the lower panel (b). The orange (black) lines correspond
to the charm quark with a constant mass of 1.5 GeV (off-shell charm quark with
a mass given by the DQPM spectral function).}
\label{qQCrossSection}
\end{figure}

Fig.~\ref{qQCrossSection} (a) shows the $q Q$ differential elastic cross sections
as a function of the scattering angle for invariant energy $\sqrt{s}=$3 and 4 GeV,
and Fig.~\ref{qQCrossSection} (b) the $qQ$ total elastic cross sections as a function
of center-of-mass energy. The temperature is taken to be $T$=0.2 GeV in the upper panel (a),
and $T$=0.2 and $T$=0.3 GeV in the lower panel (b). In these figures the orange (black)
lines correspond to the charm quark with a constant mass of 1.5 GeV
(off-shell charm quark  mass given by its DQPM spectral function). The $g Q$ elastic
cross section can be deduced from the $q Q$ process by an appropriate Casimir
color factor. Fig.~\ref{qQCrossSection} shows a sizeable difference between the
results in the two scenarios only for large scattering angles and close to the threshold.
For example, the differential cross section of off-shell charm is at most 20-25 \% larger
than that of charm  with constant mass in backward scattering (${\rm cos\theta=-1}$).
Therefore, one can conclude that introducing off-shell masses (finite width corrections)
does not change the total cross sections for heavy quark scattering on a relevant scale.
This is due to the moderate parton widths for the charm quarks considered in the DQPM model.
However, the two cross sections differ from each other below the threshold energy
where the scattering cross section of off-shell charm quarks increases with $\sqrt{s}$,
because more and more fractions of light-quark and heavy-quark spectral functions
can contribute. These have peaks at the threshold energy and then decrease
following the scattering cross sections of constant-mass charm quarks.

The scattering of charm quarks leads to an energy and momentum loss in the hot QGP
medium. The collisional energy loss of charm quark has been explicitly calculated
for on- and off-shell partons in the framework of non-perturbative QCD using the
partonic cross sections shown above in
Refs.~\cite{Berrehrah:2014kba,Berrehrah:2015ywa,Berrehrah:2013mra}.
The difference between the on- and off-shell energy losses is related to the
energy asymmetric contribution of the Breit-Wigner spectral function. We recall
that a complex propagator is used for both scenarios, which contains an
additional imaginary part proportional to the gluon width.

\begin{figure}[h]
\centerline{
\hspace{0.1 cm}\includegraphics[width=8 cm]{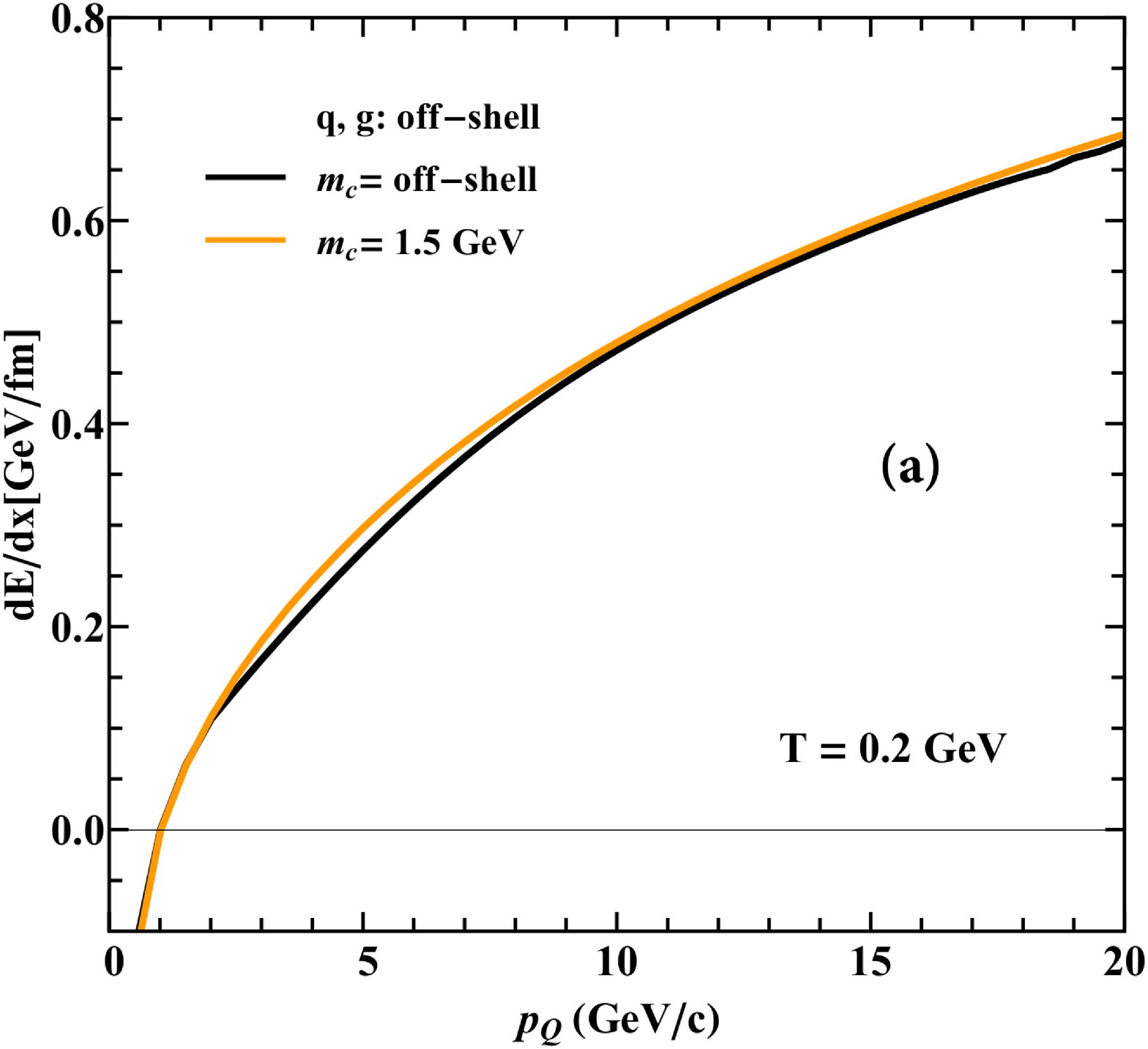}}
\centerline{
\includegraphics[width=8.2 cm]{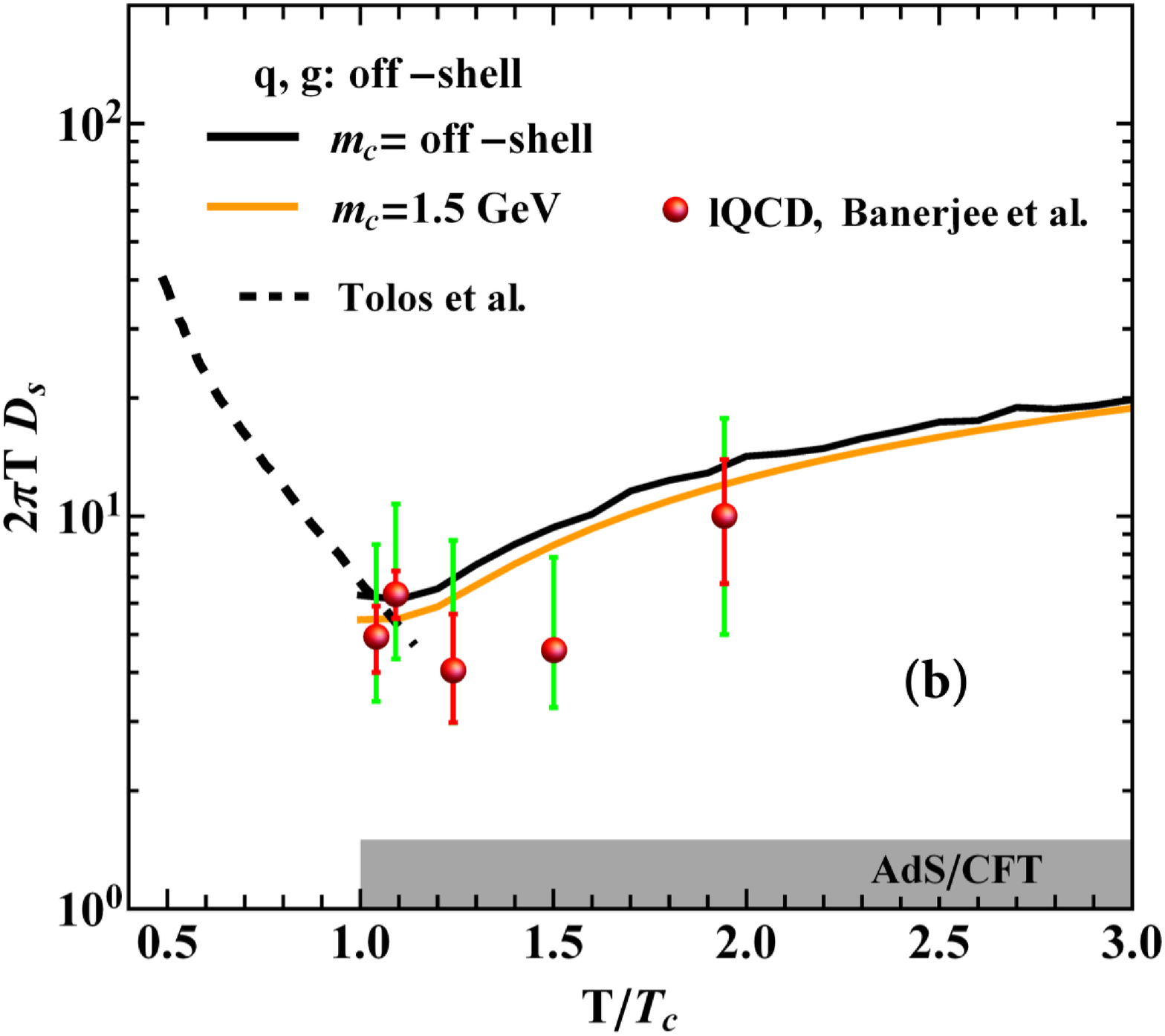}}
\caption{(Color online) Charm collisional energy loss as a function of the
charm quark momentum (a) and spatial diffusion constant $D_s$ as a function
of the medium temperature (b). The orange (black) solid lines correspond to the
charm quark with a constant mass of 1.5 GeV (off-shell charm quark with a mass
given by the DQPM spectral function). The black dashed line below $T = 180$ MeV
is the diffusion constant of $D$ mesons in hadronic matter from Ref.~\cite{Tolos:2013kva}.
The lattice QGD calculations are taken from Ref.~\cite{Banerjee:2011ra}}
\label{dEdxDs}
\end{figure}

In Fig.~\ref{dEdxDs} (a), the energy loss of charm quarks $dE/dx$ is shown
as a function of charm quark momentum for off-shell charm masses  and the charm
with a constant mass at $T = 0.2$ GeV.
The energy loss of an off-shell charm is  slightly smaller than that of a charm
with a constant mass. The figure also shows that the heavy quark gains energy
at low momentum to approach thermal equilibrium.

To validate our description of charm interactions in the QGP, the spatial diffusion
constant of charm quarks $D_s$ has been calculated on the basis of our charm scattering
cross sections~\cite{Hamza14,Berrehrah:2014ysa,Berrehrah:2014kba} and compared with
that from lQCD and that of $D$ mesons in the hadronic medium from Ref.~\cite{Tolos:2013kva}.
Fig.~\ref{dEdxDs} (b) shows a good agreement between our diffusion constants and
those from lQCD~\cite{Banerjee:2011ra} above $T_c$. For temperatures below $T_c$,
we observe that the spatial diffusion constants in hadronic and partonic matter
are smoothly connected and show a pronounced minimum around $T_c$. Finally, the
diffusion constant $D_s$ for off-shell charm is about 10 \% larger than that of
charm with a constant mass, because $D_s \propto \eta_D^{-1}$ and $\eta_D$ is
proportional to the drag coefficient which quantifies the momentum and energy losses.

The comparisons in Fig.~\ref{qQCrossSection} and \ref{dEdxDs} show that the effect
of the off-shell charm-quark mass distribution on scattering cross sections,
energy loss, and spatial diffusion constant is moderate.
We note that the most important factor, which decides the temperature-dependence
of those quantities, is the strong coupling $g(T)$ and its infrared enhancement
close to $T_c$ which is extracted from the lattice EoS.

\subsection{Hadronization}

Once the local energy density gets lower than $\sim$0.75 $\rm GeV/fm^3$,
the charm quarks may hadronize through coalescence.
In PHSD all neighboring antiquarks are candidates for the coalescence partner of the
charm quark.
From the distances in coordinate and momentum spaces between the charm quark and light
antiquark (or vice versa), the coalescence probability is given by
\begin{eqnarray}
f(\boldsymbol\rho,{\bf k}_\rho)=\frac{8g_D}{6^2}
\exp\left[-\frac{\boldsymbol\rho^2}{\delta^2}-{\bf k}_\rho^2\delta^2\right],
\label{meson}
\end{eqnarray}
where $g_D$ is the degeneracy of the $D$ meson, and
\begin{eqnarray}
\boldsymbol\rho=\frac{1}{\sqrt{2}}({\bf r}_1-{\bf r}_2),\quad{\bf k}_\rho
=\sqrt{2}~\frac{m_2{\bf k}_1-m_1{\bf k}_2}{m_1+m_2},
\label{coalescence}
\end{eqnarray}
with $m_i$, ${\bf r}_i$ and ${\bf k}_i$ being the mass, position and momentum
of the quark or antiquark $i$ in the center-of-mass frame, respectively. The
width parameter $\delta$ is related to the root-mean-square radius of the
produced $D$ meson through
\begin{eqnarray}
\langle r^2 \rangle=\frac{3}{2}\frac{m_1^2+m_2^2}{(m_1+m_2)^2}\delta^2.
\end{eqnarray}
Since this prescription gives a  larger coalescence probability at low transverse
momentum, the radius is taken to be 0.9 fm as in our previous study~\cite{Song:2015sfa}.
We also include the coalescence into highly excited states,
$D_0^*(2400)^0$, $D_1(2420)^0$, and $D_2^*(2460)^{0,\pm}$, which are assumed to
immediately decay to $D$ or $D^*$ and $\pi$ after hadronization~\cite{Song:2015sfa}.

Summing up the coalescence probabilities from all candidates, whether the charm or
anticharm quark hadronizes by coalescence or not, and which quark or antiquark among
the candidates will be the coalescence partner is decided by Monte Carlo.
If a random number is above the sum of the coalescence probabilities, it is tried again
in the next time step till the local energy density is lower than 0.4 $\rm GeV/fm^3$.
The charm or anticharm quark, which does not succeed to hadronize by
coalescence, then hadronizes through fragmentation as in p+p collisions.

\begin{figure}[h]
\centerline{
\includegraphics[width=10 cm]{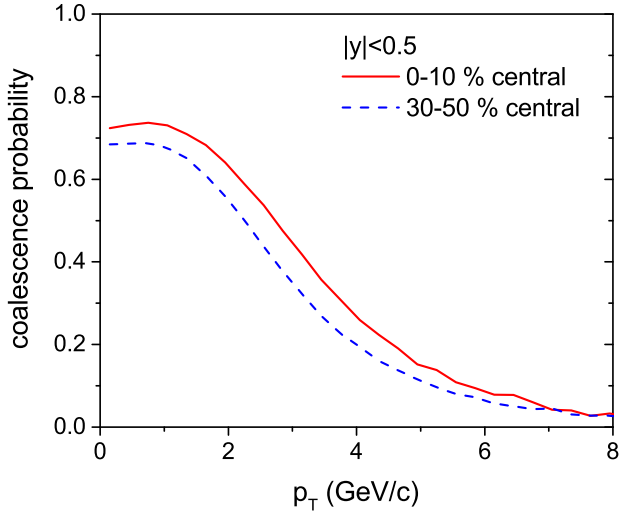}}
\caption{(Color online) Coalescence probabilities of mid-rapidity charm
($|y|<0.5$) as a function of the transverse momentum in 0-10 \% and 30-50 \%
central Pb+Pb collisions at $\sqrt{s_{\rm NN}}=$2.76 TeV.}
\label{coalpro}
\end{figure}

Fig.~\ref{coalpro} shows the coalescence probabilities of mid-rapidity charm
($|y|<0.5$) as a function of the transverse momentum in 0-10 \% and 30-50 \% central
Pb+Pb collisions at $\sqrt{s_{\rm NN}}=$2.76 TeV.
Since the charm or anticharm quark with large transverse momentum has a reduced chance
to find a coalescence partner close by in phase space, the coalescence probability
decreases with increasing transverse momentum.
Compared with the coalescence probabilities in Au+Au collisions at 200
GeV~\cite{Song:2015sfa}, the present
coalescence probabilities are slightly smaller at $\rm p_T=0$ and are
shifted to higher $\rm p_T$ essentially due to the stronger transverse flow.

Since the hadronization of charm
quarks through fragmentation is not properly suited at low
transverse momentum, several studies have forced the coalescence
probability to be 1 at $\rm p_T=$0~\cite{He:2011qa,Gossiaux:2010yx,Cao:2013ita,Oh:2009zj}.
However, it is not clear presently how to control this modeling in practice.
In the case of a hydrodynamic background, for example, the coalescence probability is precalculated in the thermalized matter and rescaled such that it is 1 at $\rm p_T=$0 as in Refs.~\cite{Gossiaux:2010yx,Cao:2013ita}.
It is no longer  1 any more after including flow effects due to the Lorentz boost.
Thus is forced to be 1 at $\rm p_T=$0 for each flow velocity in Ref.~\cite{Cao:2013ita}.
In the case of Boltzmann-type simulations, the coalescence probability depends on the system size, in other words, collision energy as well as centrality~\cite{Song:2015sfa}.
Although the coalescence probability may be rescaled at one collision energy and in a single centrality, the probability is not 1 at $\rm p_T=$0 at other collision energies or  other centralities.

We mention that it is more consistent to perform the coalescence without any additional modeling and allow a partial fragmentation at low $\rm p_T$ for the following reason:
Though the coalescence is a promising model for the hadronization of low-${\rm p_T}$ particles, it is barely applicable in p+p collisions.
For that reason, all charm quarks are assumed to be hadronized by fragmentation in p+p collisions in Sec.~\ref{initial}.
Since the fragmentation is already allowed at low-${\rm p_T}$ in p+p collision, which is the reference for the nuclear modifications in heavy-ion collision, there is no reason to prohibit fully the fragmentation in heavy-ion collisions.
Moreover, if the fragmentation is prohibited, in other words, the coalescence probability is forced to be 1 in peripheral heavy-ion collisions, it will induce a large nuclear
modification factor, although a small nuclear matter effect is expected in peripheral collisions.
Therefore, we allow a partial fragmentation at low $\rm p_T$, unless a better way of hadronization is adopted in p+p collisions.

Finally, we assume that the hadronization time of a charm quark is 0.5 $\rm fm/c$
in the rest frame, and that the charm quark does not scatter during the
hadronization. The details on the hadronization time of heavy flavor are discussed in Ref.~\cite{Song:2016lfv}.

\subsection{Hadronic interactions}

The hadronized $D$ and $D^*$ mesons then interact with hadrons in the hadron gas phase.
The cross sections for $D$ or $D^*$ scattering off pseudoscalar mesons ($\pi,~K,~\bar K,~\eta$),
baryons ($N,~\Delta$), and antibaryons ($\bar N,~\bar \Delta$) are
calculated on the basis of effective hadronic models which incorporate chiral
symmetry breaking in the light-flavor sector. The additional freedom stemming from
the coupling to heavy-flavored mesons is constrained by imposing heavy-quark spin
symmetry (HQSS)~
\cite{Tolos:2013kva,Abreu:2011ic,Abreu:2012et,GarciaRecio:2008dp,Romanets:2012hm,Garcia-Recio:2013gaa,Tolos:2013gta,Torres-Rincon:2014ffa}.
It has been shown that these cross sections have a non-trivial energy,
isospin, and flavor dependence due to the presence of resonant states close to
threshold energies with
dominant decay modes involving open-charm mesons and light hadrons~\cite{Song:2015sfa},
such as the $D_0^*(2400)$ and the $D_s^*(2317)$ in $D\pi$ and $D K$ scattering, respectively,
or the $\Lambda_c(2595)$ in $DN$ scattering, all of them dynamically generated in these approaches.
The cross sections for scattering off other light hadrons (such as the vector mesons from the octet),
which are not calculated above, are taken to be 10 mb and independent of the collision energy.

\section{Results}\label{results}

The medium effect on charm production in relativistic heavy-ion collisions
is expressed in term of the ratio $R_{\rm AA}$, which is defined as
\begin{eqnarray}
R_{\rm AA}({\rm p_T})\equiv\frac{dN_D^{\rm Pb+Pb}/d{\rm p_T}}{N_{\rm binary}^{\rm Pb+Pb}\times dN_D^{\rm p+p}/d{\rm p_T}},
\label{raa}
\end{eqnarray}
where $N_D^{\rm Pb+Pb}$ and $N_D^{\rm p+p}$ are, respectively, the numbers of
$D$ mesons produced in Pb+Pb collisions and in p+p collisions, and
$N_{\rm binary}^{\rm Pb+Pb}$ is the number of binary nucleon-nucleon
collisions in Pb+Pb collisions for the centrality class considered.
$R_{\rm AA}$ larger (smaller) than 1.0 indicates that nuclear matter enhances
(suppresses) charm production in relativistic heavy-ion collisions.

\begin{figure}[h]
\centerline{
\includegraphics[width=10 cm]{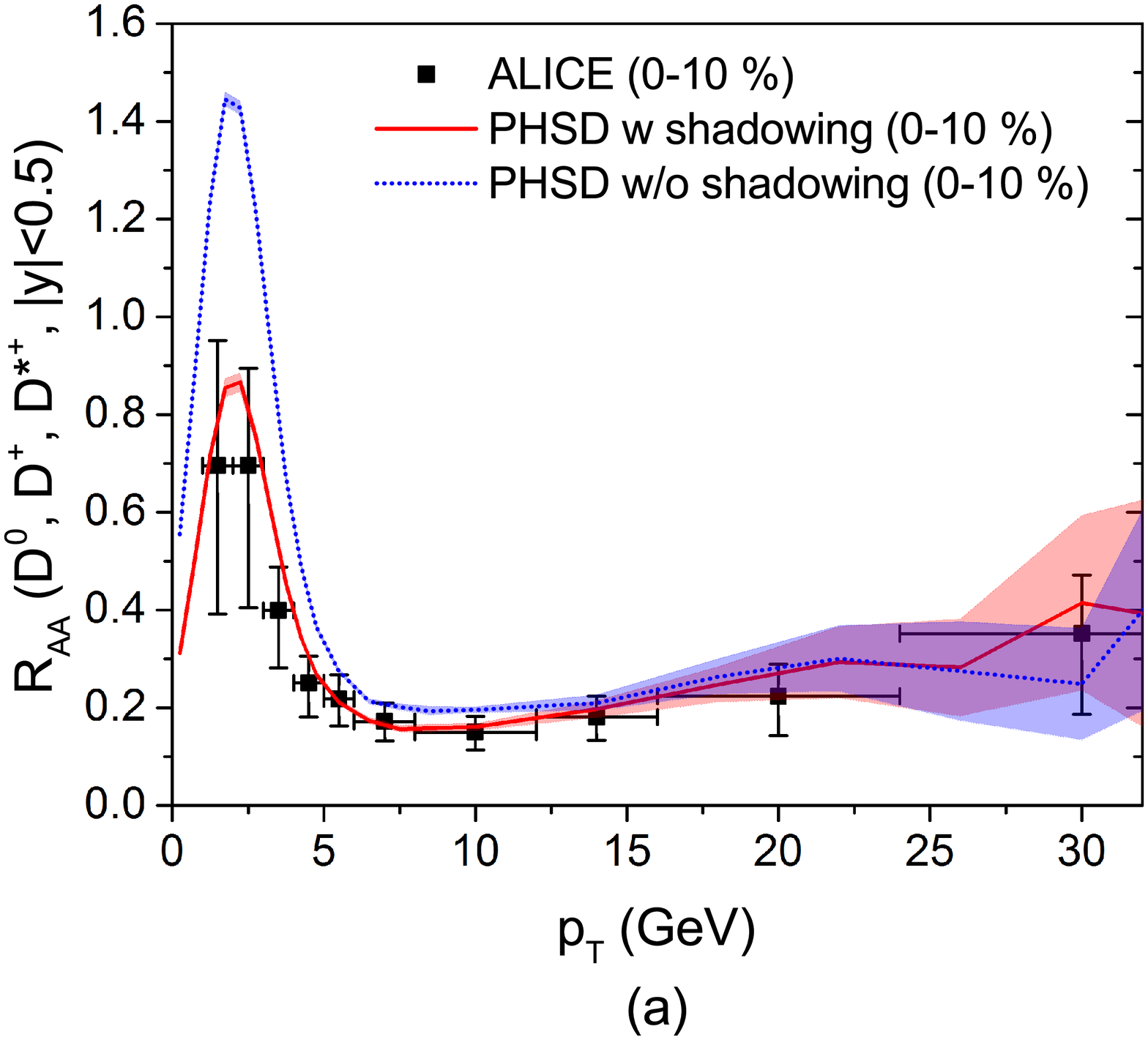}}
\centerline{
\includegraphics[width=10 cm]{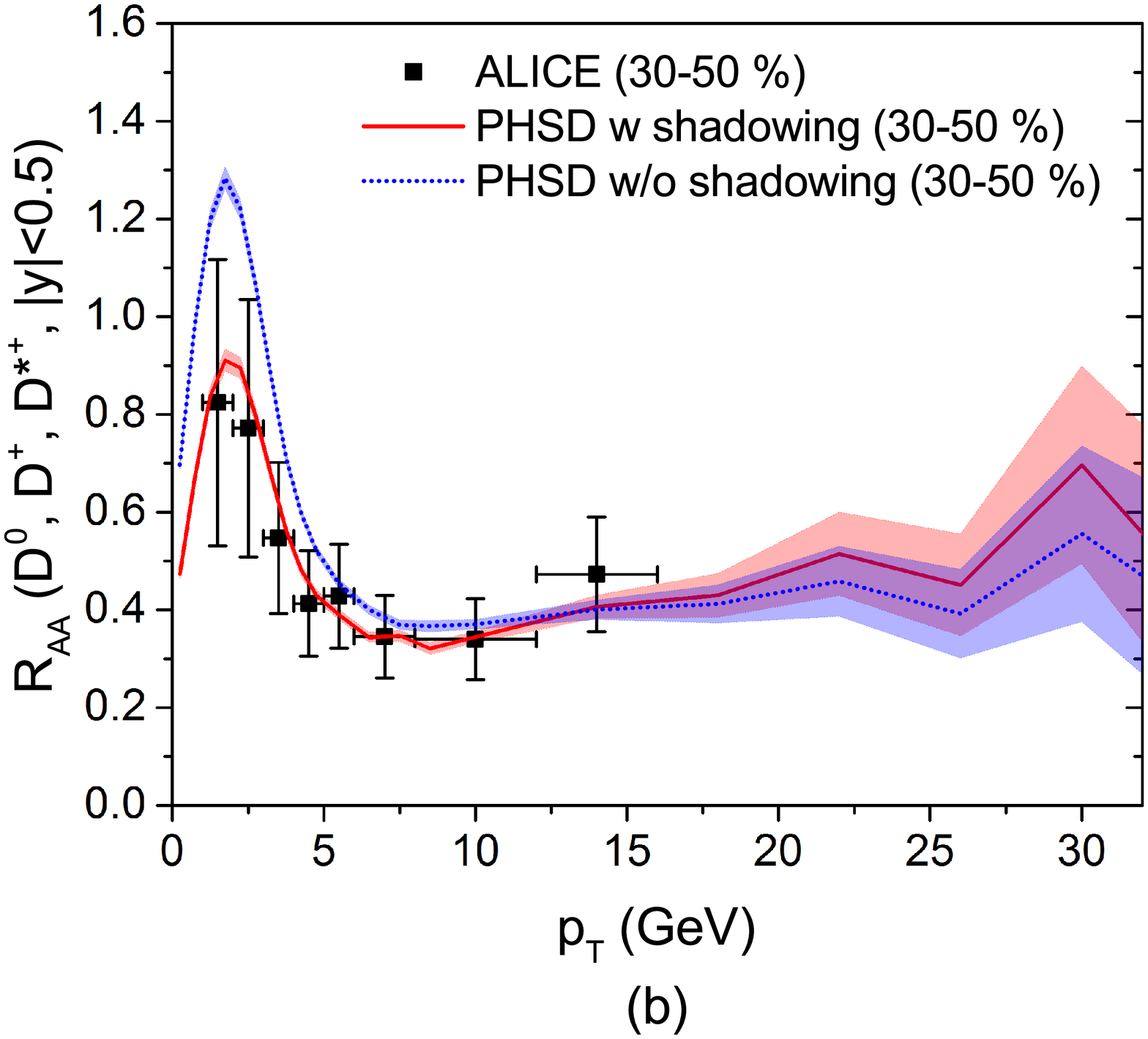}}
\caption{(Color online) The ratio  $R_{\rm AA}$ of $D^0,~D^+$, and $D^{*+}$ mesons
within $|y|<0.5$ as a function of ${\rm p_T}$ in 0-10 \% (a) and 30-50 \% (b)
central Pb+Pb collisions at $\sqrt{s_{\rm NN}}=$2.76 TeV compared with the
experimental data from the ALICE collaboration~\cite{Adam:2015sza}. The solid and
dotted lines are, respectively, $R_{\rm AA}$ with and without (anti-)shadowing.
The charm quark mass is taken to be 1.5 GeV.}
\label{raa0}
\end{figure}

Fig.~\ref{raa0} shows the ratio $R_{\rm AA}$ of $D^0,~D^+$, and $D^{*+}$ mesons
within the rapidity range $|y|<0.5$ as a function of ${\rm p_T}$ in 0-10 \% and
30-50 \% central Pb+Pb collisions at $\sqrt{s_{\rm NN}}=$2.76 TeV.
Here the charm quark mass is taken to be 1.5 GeV, independent of temperature.
The solid and dotted lines are, respectively, the $R_{\rm AA}$ of $D$ mesons with
and without (anti-)shadowing.
We can see that the $R_{\rm AA}$ of $D$ mesons decreases especially at small
transverse momentum due to shadowing, which is consistent with Fig.~\ref{shadow0}.
When including  shadowing our results are in a good agreement with the experimental
data from the ALICE collaboration~\cite{Adam:2015sza}.
Since the radiative energy loss is not yet included in our study, it seems that the
latter is not significant for transverse momenta up to 15 GeV/c.
At higher transverse momenta our statistics is too low to allow
for a solid answer. We speculate that the dominance of partonic scattering is due to the
fact that in PHSD the scattering partners of the charm quarks are massive partons.

The elliptic flow is generated in non-central heavy-ion collisions due to
asymmetric pressure gradients in the transverse plane, and expressed in term of
the coefficient $v_2$ defined as
\begin{eqnarray}
v_2({\rm p_T})\equiv\frac{\int d\phi \cos2\phi (dN_D^{\rm Pb+Pb}/d{\rm p_T}d\phi)}{dN_D^{\rm Pb+Pb}/d{\rm p_T}},
\end{eqnarray}
where $\phi$ is the azimuthal angle of the $D$ meson in momentum space.

\begin{figure}[h]
\centerline{
\includegraphics[width=10 cm]{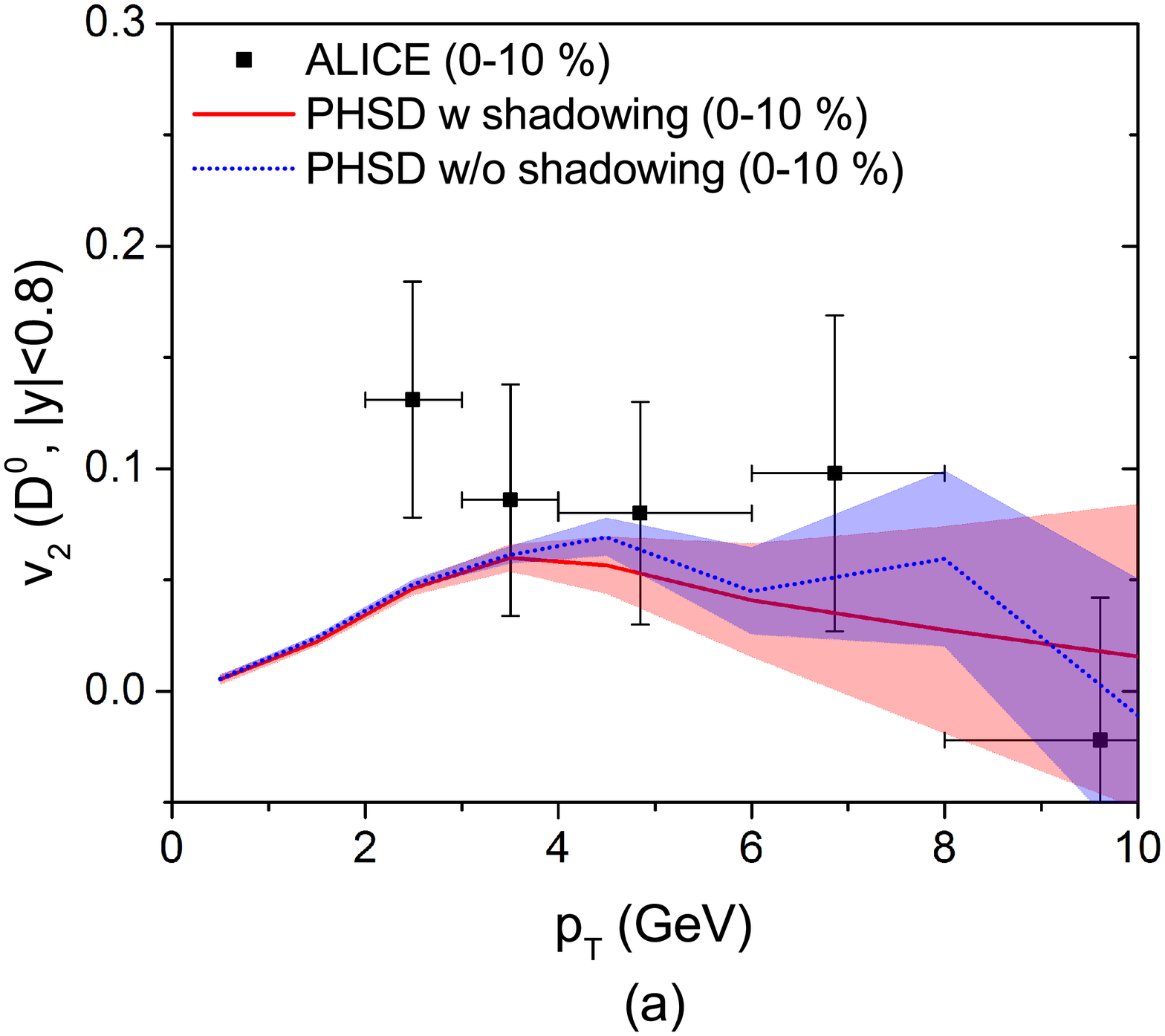}}
\centerline{
\includegraphics[width=10 cm]{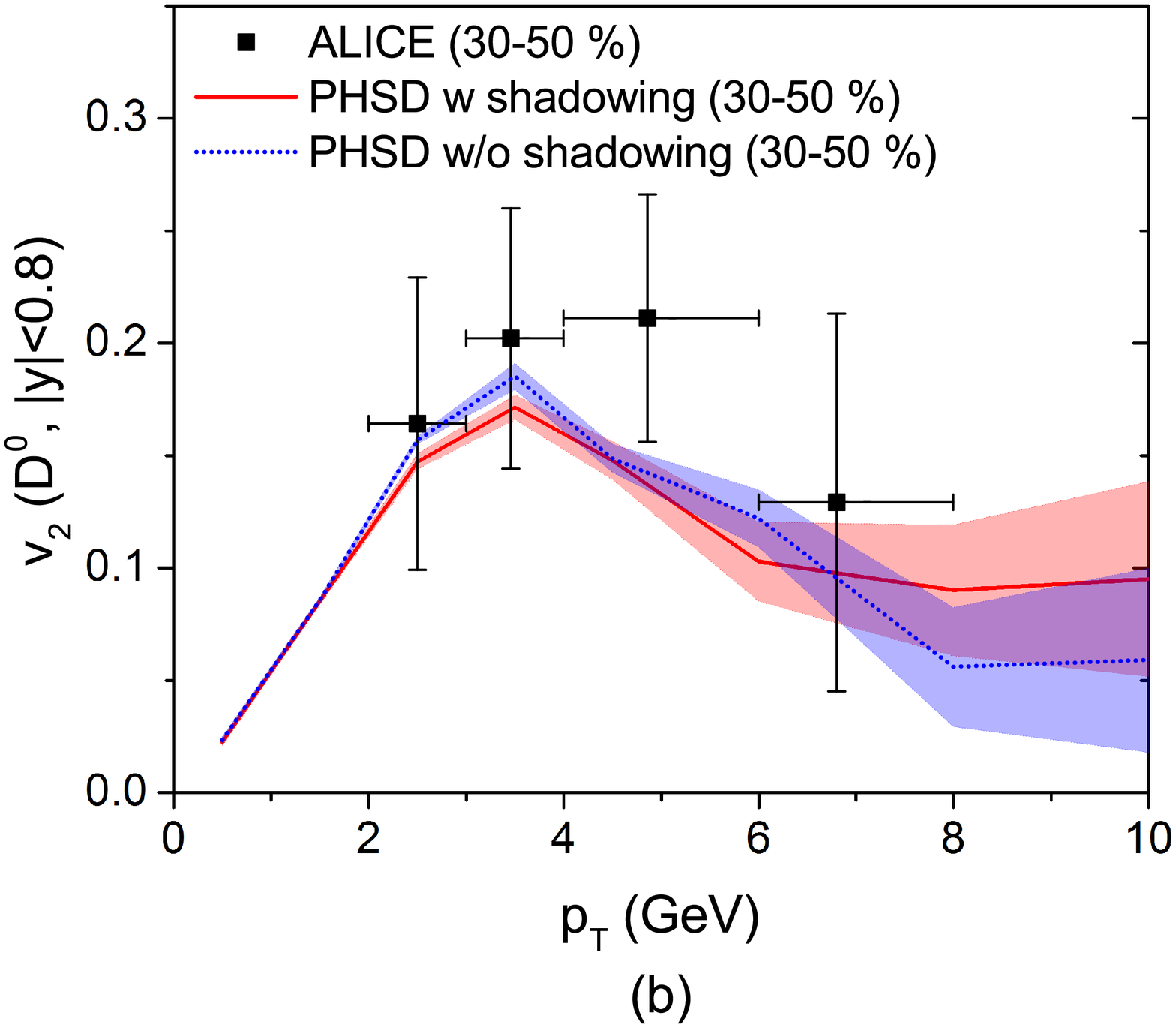}}
\caption{(Color online) The elliptic flow $v_2$ of $D^0$ mesons within $|y|<0.8$
in 0-10 \% (a) and 30-50 \% (b) central Pb+Pb collisions at $\sqrt{s_{\rm NN}}=$2.76
TeV compared with the experimental data from the ALICE
collaboration~\cite{Abelev:2014ipa}. The solid and dotted lines are, respectively,
$v_2$ with and without (anti-)shadowing. The charm quark mass is taken to be 1.5 GeV.}
\label{v2}
\end{figure}

Fig.~\ref{v2} shows the elliptic flows $v_2$  of $D^0$ mesons within the rapidity
range $|y|<0.8$ in 0-10 \% and 30-50 \% central Pb+Pb collisions at
$\sqrt{s_{\rm NN}}=$2.76 TeV.
The solid and dotted lines are, respectively, the results with and without
(anti-)shadowing. Again the charm quark mass is taken to be 1.5 GeV.
We can see that the PHSD results reproduce the experimental data
from the ALICE collaboration~\cite{Abelev:2014ipa}.
The (anti-)shadowing effect slightly decreases the elliptic flows, because it
reduces the production of low-$\rm p_T$ charm which more easily follows the bulk flow.
It has been a challenge for theoretical models to reproduce the experimental data and to explain simultaneously the large energy loss of charm quarks ($R_{\rm AA}$) and the strong collectivity ($v_2$)~\cite{Andronic:2015wma}.
According to a recent study~\cite{Das:2015ana}, both $R_{\rm AA}$ and $v_2$ are well reproduced if the drag coefficient for charm quarks is large close to the critical temperature.
For our drag coefficient this is the case, since the strong coupling fitted to the lattice EoS rapidly increases near the critical temperature.

We point out that -- for the same scattering cross sections and hadronization
processes --  our results are in good agreement with the experimental data
in Au+Au collisions at $\sqrt{s_{\rm NN}}=$200 GeV as well as in Pb+Pb collisions
at $\sqrt{s_{\rm NN}}=$2.76 TeV, which spans more than one order of magnitude
in collision energy $\sqrt{s}$.
Considering that the PHSD transport approach provides good bulk dynamics
from SPS to LHC energies \cite{PPNP16}, our
description of charm production and charm interactions in relativistic
heavy-ion collisions appears consistent.

Finally, we study the effect of off-shell charm on the charm production
and propagation in relativistic heavy-ion collisions.
We recall that the spectral function with a finite width has both time-like
and space-like parts~\cite{Cassing:2009vt}.
The time-like parton propagates in space-time while the space-like one is
interpreted as a virtual parton which mediates the interaction between time-like partons.
The latter contributes to the potential energy density in the QGP~\cite{Cassing:2009vt}.
The potential energy density is then separated into scalar and (time-component)
vector parts by using the energy density and pressure of the lattice
EoS~\cite{Cassing:2009vt}. At RHIC and LHC energies the scalar
potential dominates by far due to approximately equal densities of
quarks and antiquarks.
The scalar potential energy density increases with increasing temperature
(or scalar parton density) except
near $T_c$~\cite{Cassing:2009vt}, and thus gives a repulsive force on partons
in relativistic heavy-ion collisions (except close to $T_c$ or during hadronization).
The repulsive force for gluons is roughly twice stronger than that for quarks or antiquarks,
because one gluon is roughly equivalent to a quark and antiquark pair.
It is presently not clear how much a charm quark is affected by scalar partonic
forces.

\begin{figure}[h]
\centerline{
\includegraphics[width=10 cm]{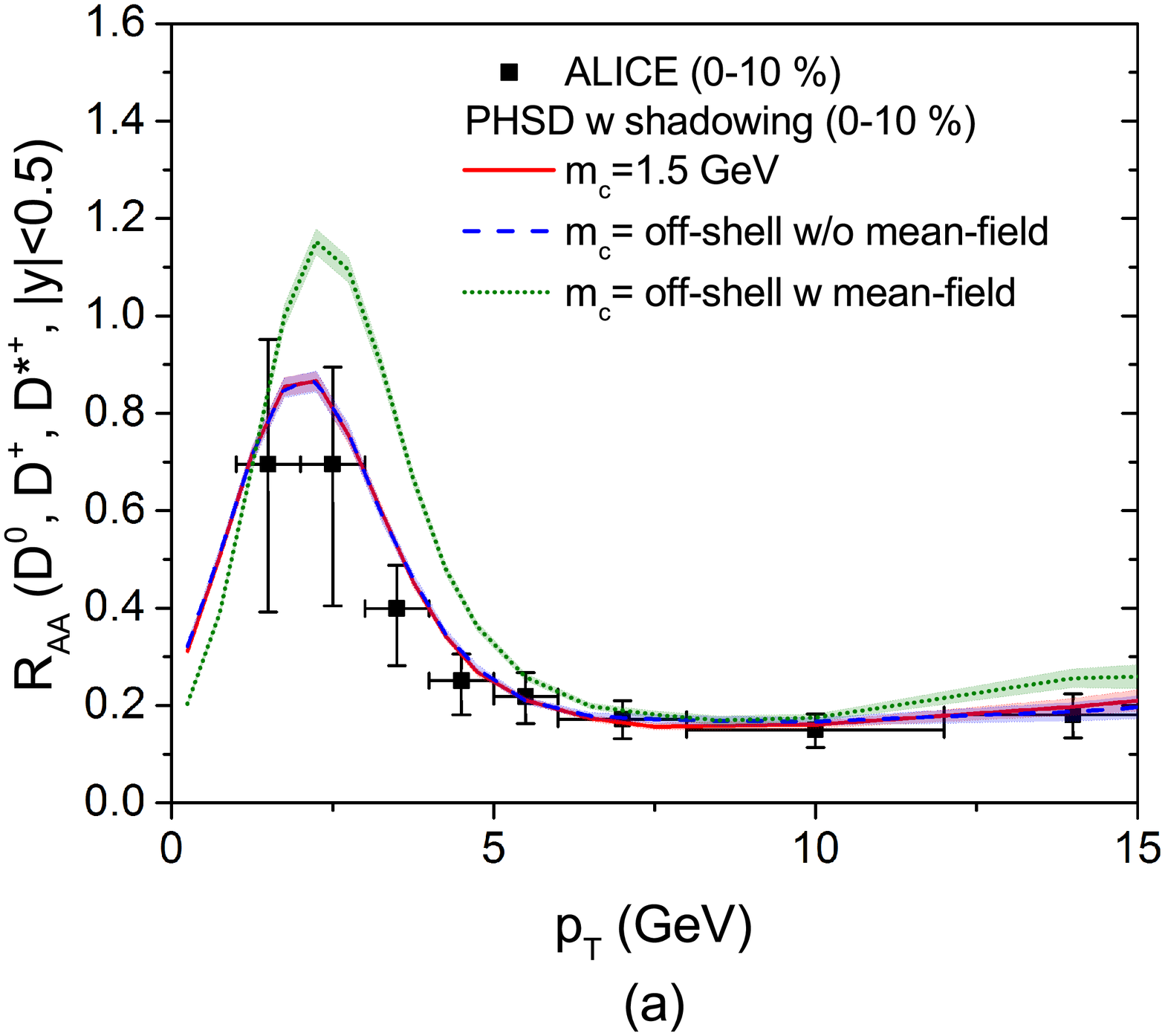}}
\centerline{
\includegraphics[width=10 cm]{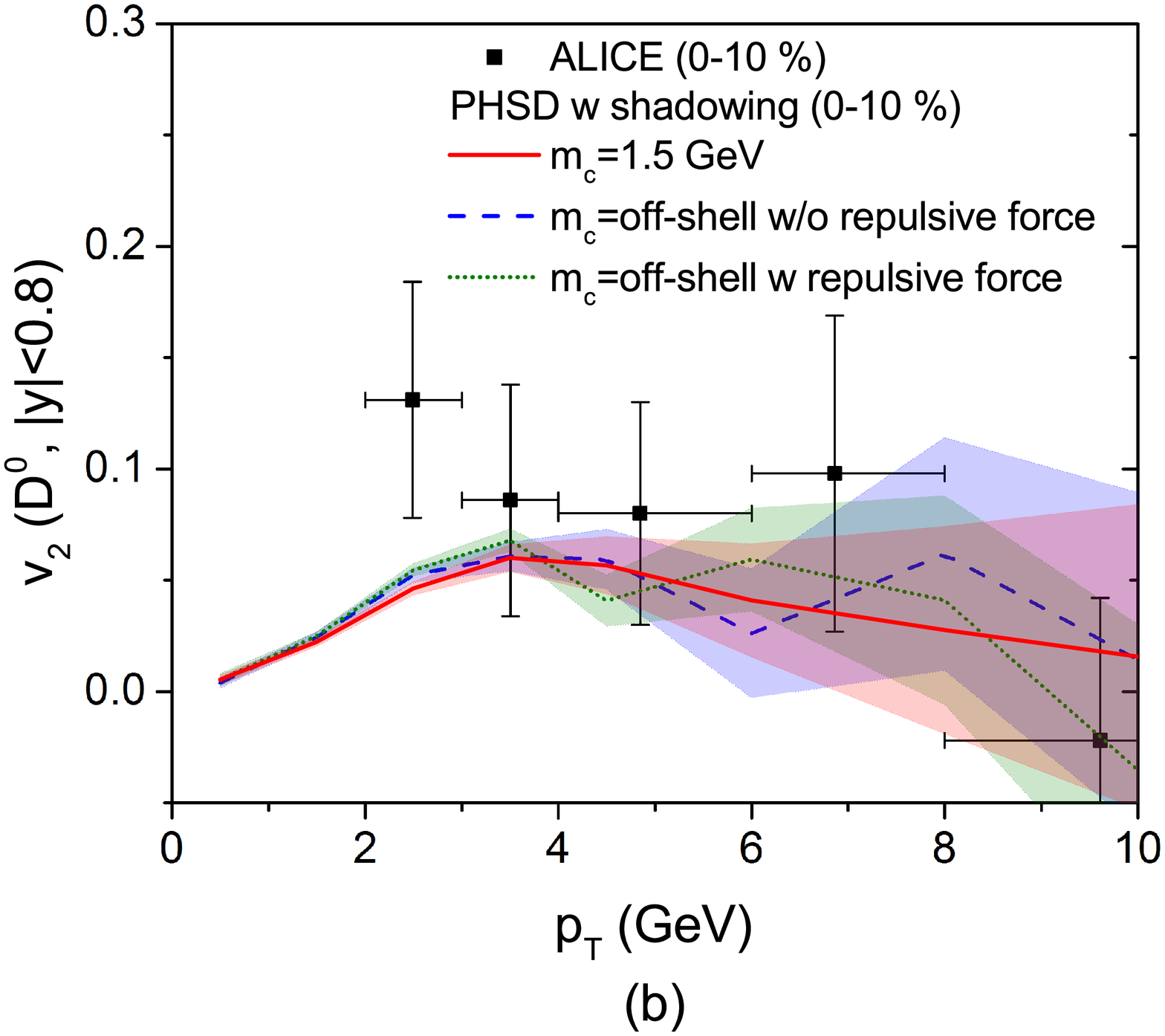}}
\caption{(Color online) The ratio $R_{\rm AA}$ of $D^0,~D^+$, and $D^{*+}$ mesons
within $|y|<0.5$ (a) and the elliptic flow $v_2$ of $D^0$ mesons within
$|y|<0.8$ (b) as a function of ${\rm p_T}$ in 0-10 \% central Pb+Pb collisions
at $\sqrt{s_{\rm NN}}=$2.76 TeV compared with the experimental data from
the ALICE collaboration~\cite{Adam:2015sza}. The solid, dashed and dotted lines are,
respectively, for charm quarks with the mass of 1.5 GeV, and for off-shell charm
quarks without and with the repulsive force originated from the scalar potential
energy density for light quarks. (Anti-)shadowing is included in all cases.}
\label{off-shell}
\end{figure}

Fig.~\ref{off-shell} shows the ratio $R_{\rm AA}$ of $D^0,~D^+$, and $D^{*+}$ mesons
within $|y|<0.5$ (a) and the elliptic flow $v_2$ of $D^0$ mesons within $|y|<0.8$ (b)
as a function of ${\rm p_T}$ in 0-10 \% central Pb+Pb collisions at
$\sqrt{s_{\rm NN}}=$2.76 TeV.
The solid, dashed and dotted lines are, respectively, for charm quarks
with the mass of 1.5 GeV, and for off-shell charm quarks without and with the
repulsive force originating from the scalar potential energy density.
We note that (Anti-)shadowing is included in all cases.
The comparison between the solid and dashed lines shows that the effect of
off-shell charm without repulsive force on $R_{\rm AA}$ and $v_2$ is small,
which is expected from Figs.~\ref{qQCrossSection} and \ref{dEdxDs}.
When including the repulsive force, the peak of the ratio $R_{\rm AA}$ is shifted
to larger transverse momentum, as shown by the dotted lines.
However, the comparison with the experimental data from the ALICE collaboration
favors a weaker repulsive force for off-shell charm quarks compared to light quarks.

\begin{figure}[h]
\centerline{
\includegraphics[width=10 cm]{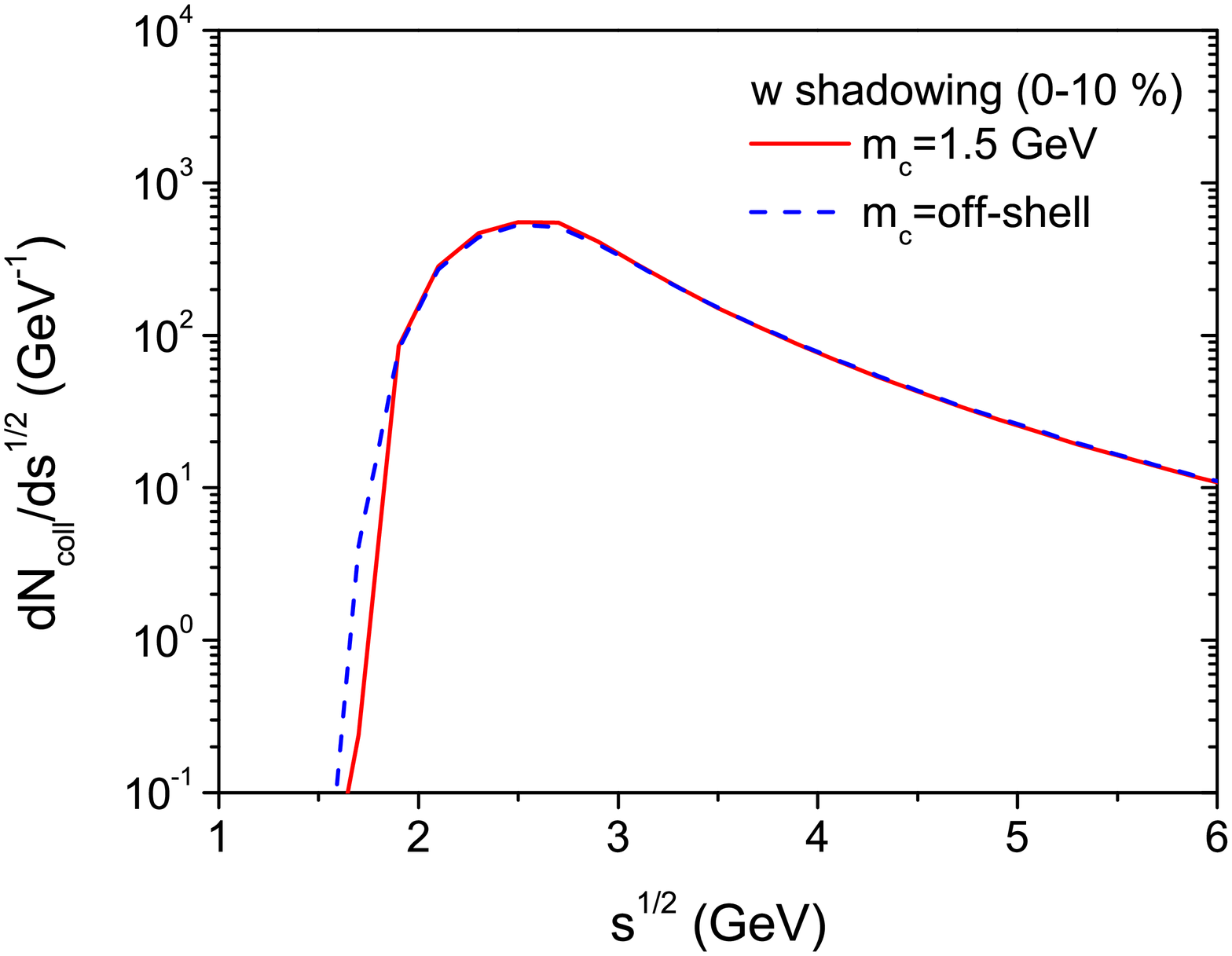}}
\caption{(Color online) the distributions of charm quark scattering in QGP as functions of collision energy for the constant mass and the off-shell mass of charm quark in 0-10 \% central Pb+Pb collisions at 2.76 TeV with the shadowing effect included.}
\label{scatterq}
\end{figure}

Fig.~\ref{scatterq} shows the distributions of charm quark scattering in the QGP as a function of collision energy for the constant mass and the off-shell mass of charm quarks in 0-10 \% central Pb+Pb collisions at 2.76 TeV (with the shadowing effect included).
The scattering distributions are similar to each other except at very low $\sqrt{s}$ where the number of scatterings is larger in the case of off-shell charm.
As shown in Fig.~\ref{qQCrossSection} (b), off-shell charm has a sizeable scattering cross sections even below $\sqrt{s}=1.5$ GeV. However, it induces only little additional scatterings because the mass of the charm quark -- involved in such scatterings -- is much smaller than the pole position of the charm spectral function and occurs with low probability.
Considering that the number of produced charm quarks is about 84 and the total number of charm quark scattering 720 in 0-10 \% centrality, one charm quark experiences on average 9 elastic scatterings before hadronization in central collisions.

\section{Summary}\label{summary}

We have studied charm production in Pb+Pb collisions at $\sqrt{s_{\rm NN}}=$2.76 TeV
in the PHSD transport approach in continuation of our calculations for Au+Au collsions
at the top RHIC energy~\cite{Song:2015sfa}.
The initial charm quark pairs are produced through nucleon-nucleon binary collisions
by using the PYTHIA event generator.
The (anti-)shadowing, which is a modification of the parton distributions in a nucleus,
has been implemented in the PHSD by using the EPS09 package.
We have found that  (anti-)shadowing reduces the charm quark production preferentially
at low transverse momentum and at mid-rapidity.

The produced charm and anticharm quarks interact in the QGP with quarks and gluons
whose masses are given by spectral functions with pole positions and widths
being fitted to the lattice QCD equation-of-state (EoS) \cite{PPNP16}.
Since the contribution from charm quarks to the energy density and pressure of the QGP
is small, the spectral function of a charm quark cannot be constrained by the lattice EoS.
Therefore, we have studied two different scenarios:
In the first case the charm quark mass is 1.5 GeV independent of temperature, i.e. the
charm quark spectral function is a delta function peaked at 1.5 GeV.
In the second case, the pole position and width of the charm spectral function
are assumed to have the same temperature-dependence as those of light quarks except
that the pole position is shifted by 1 GeV from the pole position of a light quark.
We have pointed out that our partonic cross sections reproduce the spatial diffusion
constant of heavy quarks from lQCD and smoothly join with that of $D$ mesons
in a hadron gas close to the critical temperature $T_c$.

Once the local energy density gets lower than the critical value of the
crossover transition
in the expansion of the system, the charm quark is hadronized into a $D$ meson
(or its excited states) through either coalescence or fragmentation.
The coalescence probability of a charm quark is calculated from the light antiquarks
close in coordinate and momentum space. Accordingly, a
charm quark with a small transverse momentum has a large coalescence probability,
while the one with a large transverse momentum has a small probability.
If coalescence is rejected in the Monte-Carlo method, the charm quark is
hadronized through fragmentation as in p+p reactions.
The essential difference between coalescence and fragmentation is that a charm
quark gains transverse momentum in the first case while it loses transverse momentum
in the second one.

The hadronized $D$ mesons then interact with hadrons in the hadron gas phase.
The cross sections for $D$ or $D^*$ meson scattering off light pseudoscalar mesons,
light baryons and antibaryons are calculated in an effective Lagrangian approach
with heavy-quark spin symmetry.
After several hundred fm/c, depending on collision centrality, the $D$ mesons freeze
out, and are analyzed in momentum for the comparison with experimental data.

As a result from PHSD, both the $R_{\rm AA}$ and the elliptic flow $v_2$ of $D$ mesons from the
ALICE collaboration are reasonably reproduced.
This supports the validity and consistency of the PHSD approach for charm production
and propagation in relativistic heavy-ion collisions in connection with our previous
study at the top RHIC energy within the same approach~\cite{Song:2015sfa}.

Furthermore, we have found that the shadowing effect suppresses charm production
preferentially at small transverse momentum and  mid-rapidity, and it helps the
PHSD to reproduce the ratio $R_{\rm AA}$ of $D$ mesons from the ALICE collaboration.
The shadowing also slightly decreases the elliptic flow of $D$ mesons
because it suppresses the production of charm quarks with small transverse momentum
which more easily follow the bulk flow.

Finally, we have studied the effect of off-shell charm on the charm production
and propagation in relativistic heavy-ion collisions.
{It has been found that the scattering cross sections are only moderately
affected by the off-shell charm quarks, but the repulsive force -- generated
by a scalar potential energy -- shifts the peak of $R_{\rm AA}$ of $D$ mesons
to higher transverse momentum when assuming the same strength as for the light quarks.
The comparison with the experimental data on the ratio $R_{\rm AA}$ of $D$ mesons
with the actual data from the ALICE collaboration supports a weaker repulsive force
for off-shell charm quarks than for the off-shell light quarks.}

PHSD is a microscopic transport model that allows for the detailed study of the charm dynamics in hot and dense QCD matter within a non-equilibrium transport setting.
The specific non-equilibrium features of the PHSD model sets it apart from other theoretical approaches which assume thermal equilibrium.
Our future study, which compares and contrasts the results from PHSD with those from other approaches, should give an insight into the effect of non-equilibrium matter on the production and interactions of heavy flavor in relativistic heavy-ion collisions.

\section*{Acknowledgements}
The authors acknowledge inspiring discussions with J. Aichelin, P. B.
Gossiaux, C. M. Ko, O. Linnyk, J. Torres-Rincon, L. Tolos, V. Ozvenchuk, and R.
Vogt.
This work was supported by DFG under contract BR 4000/3-1 and
by the LOEWE center "HIC for FAIR". The computational resources have been
provided by the LOEWE-CSC.

\end{document}